\documentclass[iop]{emulateapj}
\usepackage{pdfpages}
\usepackage{epsfig}
\usepackage{epstopdf}

\usepackage[english]{babel}

\RequirePackage{color}
\def\H2{H$_{2}$}

\def\roH2{$\rho_{\textrm{H}_2}$}

\def\MH2{M$_{\textrm{H}_2}$}

\begin{document}

\title{On the effective oxygen yield in the disks of spiral galaxies}
\shorttitle{On the effective oxygen yield}
\shortauthors{Zasov, Saburova, Abramova }

\author{A. Zasov\altaffilmark{1,2}} 
\email{zasov@sai.msu.ru}
\author{A. Saburova\altaffilmark{1}}
\email{saburovaann@gmail.com}
\author{O. Abramova\altaffilmark{1}}

\affil{$^1$ -- Sternberg Astronomical Institute, Moscow M.V. Lomonosov State
University}

\affil{$^2$ -- Department of Physics, Moscow M.V. Lomonosov State
University}

\begin{abstract}

The factors influencing chemical evolution of galaxies are  poorly understood. Both gas inflow and gas outflow reduce a gas-phase  abundance of heavy elements (metallicity) whereas the ongoing  star formation continuously increases it. To exclude the stellar nucleosynthesis from consideration, we analyze for the sample of 14 spiral galaxies the radial distribution of the effective yield of oxygen $y_{eff}$, which would be identical to the true stellar yield (per stellar generation) $y_o$ if the evolution followed  the closed box model.  As the initial data for gas-phase abundance we used the O/H radial profiles from \citet{Moustakas2010}, based on two different calibrations (\citealt{PT2005} (PT2005) and \citealt{KK2004} (KK2004) methods). In most of galaxies with the PT2005 calibration, which we consider as a preferred one, the yield $y_{eff}$  in the main disk ($R \ge 0.2~R_{25}$, where $R_{25}$ is the optical radius) increases  with radius, remaining lower than the empirically found true stellar yield $y_o$. This may indicate the inflow of low-enriched gas predominantly to the inner disk regions, which reduces $y_{eff}$.   
We show that the maximal values of the effective yield in the main disks of galaxies, $y_{eff,max}$, anti-correlate with the total mass of galaxies and with the mass of their dark halo enclosed within  $R_{25}$. It allows to propose the higher role of gas accretion for galaxies with massive halos.  We also found that the radial gradient of oxygen abundance normalized to $R_{25}$ has a tendency to be shallower in the systems with lower dark halo to stellar mass ratio within the optical radius, which, if confirmed, gives evidences of the effective radial mixing of gas in galaxies with the relatively light dark matter halo.
 
	\end{abstract}

\section{Introduction}

Gas-phase metal abundance (gas metallicity) in the disks of galaxies depends on their star formation history, and generally grows with time as the result of the enrichment of interstellar gas by the products of stellar nucleosynthesis.  The observed relations between  gas metallicity and stellar disk properties, such as  stellar mass-metallicity relation (\citealt{Pilyugin2013}; \citealt{Zahid2014}),  or local metallicity versus local disk surface density relation  (\citealt{Moran2012}; \citealt{Rosales-Ortega2012}; \citealt{Sanchez2013}; \citealt{Pilyugin2014b})   give evidence that the chemical evolution is mostly a product of internal processes rather than the environment, although the gas exchange between  disk and a halo or intergalactic medium  also plays a significant role, especially at the early stage of evolution. 

Following \cite{Ascasibar2015}, we proceed  from the assumption that there are two parameters directly obtained from observations, which  describe the evolutionary stage of a starforming galaxy: a gas mass fraction $\mu = M_g/(M_g+M_*)$, where the $\textit{g}$ and $ \textit{*}$ indices  are related to gas and stellar population masses respectively, and gas metallicity, usually presented by the oxygen abundance  12+$\log(O/H)$, which is a product of short-lived stars. In general case, both parameters vary along the galactocentric radius $R$ and change in time. A continuous star formation leads to the monotonous decreasing of $ \mu$ with the timescale of several Gyr, accompanied by the corresponding growth of $O/H$,  however the situation becomes more  complex in the presence of gas accretion or the gas losses.  

The existence of proceeding gas accretion onto galaxies which dilutes the interstellar gas, and gas outflow from the disks of galaxies is supported by many observational evidences (see f.e. the review by \citealt{Almeida2014}). The efficiency of gas inflow/outflow processes and its role in chemical evolution  has been discussed and modelled in many papers (see f.e. \citealt{Pilyugin2007};  \citealt{Dalcanton2007}; \citealt{Spitoni2010}; \citealt{Zahid2014}; \citealt{Ascasibar2015};  \citealt{Kudritzki2015}; \citealt{Lu2015}). However the results are still contradictory and model-dependent. In particular, there remains a degeneracy between the influence of gas inflow and the outflow of enriched gas onto chemical evolution, because both of them tend to decrease a gas-phase metallicity. Note, however, that the accretion and wind can in principle be disentangled (see \citealt{Kudritzki2015}). Although the results remain model dependent \cite{Kudritzki2015} demonstrated that the role of gas inflow/outflow may be different for different spiral galaxies.
 
 The accretion rate, as well as the intensity of galactic wind or radial gas migration in the disk plane  depends on the  deepness of potential well, provided by the extended dark halo. In particular, the baryonic mass growth of a disk due to accretion may be less efficient both for low mass halos and for very massive halos $M_{halo} \ge 10^{12} M\odot$ (\citealt{Bouche2010}). The latter is caused by the shock heating of cool gas flows penetrating  into a massive halo from the intergalactic spare, so that it requires a long time for the hot halo gas to cool and to fall onto a disk (see the discussion in  \citealt{Almeida2014} and \citealt{vandeVoort2011}).  In turn, the massive halo reduces or prevents the gas losses of a galaxy, so that the enriched gas which leaves a disk as the result of stellar feedback may replenish the gas in the halo and return back to the disk much later. The greater the mass of a halo, the hotter and thinner its gas, and the less efficiently it cools before settling down onto a disk. However the relation between $M_{halo}$ and the efficiency of gas accretion and gas losses depends on many factors and remains badly known, although some restrictions may be done for the net accretion rate
and the net mass outflow rate based on the comparison of observational data with the model predictions under some natural assumptions (\citealt{Lu2015}).

There are big uncertainties in modelling of the mixing of metals  with the medium, so the results given by different models of chemical evolution, even being compatible with observational data, remain model dependent and rather contradictory concerning the role of gas accretion and galactic winds in concrete galaxies. In addition, a true yield of metals in the stellar nucleosynthesis is also poorly known. Moreover, there is no agreement between the different methods of evaluation of gas-phase metallicity, so that the estimates, based on the radiation balance equation of H II regions, are systematically higher than those obtained by semi-empiric methods: the difference may be as high as a factor of 5 between the O/H ratio estimates (\citealt{Kewley2008}).  

  In this work we aim to study the  properties of gas-phase evolution of metal content in different  spiral galaxies  avoiding the using of  sophisticated evolution models. To exclude from consideration a history of gas consumption in the formation of stars we consider below the effective yield of oxygen $y_{eff}$ which by definition is equal to the true stellar yield of oxygen $y_o$ per stellar generation if the model of closed box evolution is valid. In this simple model the gas-phase metallicity is governed by star formation without the inflow or outflow of gas,  gas enrichment is instantaneous and gas is well mixed. The oxygen yield $y_o$ may be calculated from theoretical models of stellar evolution (see f.e. \citealt{Vincenzo2015}), or found by empirical way (\citealt{Pilyugin2007}). In the closed box model the oxygen mass fraction Z$\approx12 (O/H)$ is linked with the gas-to-stellar mass ratio (or, locally,  gas surface density fraction $\mu$) by simple relation (\citealt{Searle1972}; \citealt{Edmunds1990}; \citealt{Belfiore2015}) 
\begin{equation}
Z = \frac{y_o}{1-r}\cdot \ln(1/\mu)
\end{equation}
where $r$ is the total mass fraction (including both processed and unprocessed material) returned back into the interstellar medium as the result of stellar evolution. Actually, the closed box model is too primitive for describing the evolution of stellar-gaseous disks of galaxies. 

The effective yield is defined as
\begin{equation}
y_{eff} = \frac{Z}{\ln(1/\mu)}
\end{equation}
Its value does not depend on star formation efficiency or the history of star formation, differing from the true stellar yield $y_o$ by the nominator $(1-r)$ in the closed box model. In the general case $y_{eff}$ may strongly differ from $y_o$. A  comparison of these two yields may give a valuable information for concrete galaxies about the factors influencing the chemical evolution of gas besides the stellar nucleosynthesis. For example,  \citealt{Kudritzki2015} showed with their model that the high effective yield of galaxies may be explained by relatively low rates of accretion and winds. In general case, $y_{eff}$ may change drastically along the radius. In this paper we focus on the radial trend of the effective yield.

\section{Radial profiles of the effective oxygen yield}\label{rad} 

Both the inflow of low-metal gas and the outflow of the enriched gas from the disk results in the reducing of $y_{eff}$ (\citealt{Dalcanton2007}). It is only the weak mixing of the enriched and non-enriched gas or radial migration of the enriched gas from the inner regions outwards that can lead to an increase of $y_{eff}$ in the disk periphery. The latter manifests itself as the flattening of the (O/H) gradient at large radial distances R, mostly observed in galaxies which experience or have experienced a strong interaction (\citealt{Werk2011}; \citealt{Zasov2015}). A comparison of $y_{eff}$ and $y_{o}$ has been used by different authors to clarify the degree of difference of evolution from the closed-box model and to specify the processes which are responsible for the metal abundance besides the star formation and gas consumption (see f.e, \citealt{Zahid2014}; \citealt{Kudritzki2015}; \citealt{Ascasibar2015}). 

Below we present the radial profiles of the effective oxygen yield $y_{eff}$ for 14 spiral galaxies 
(see Table \ref{tab} for the names of the galaxies, the adopted distances in Mpc, morphological  types and B-band absolute magnitudes), obtained for the closed box model: 
 \begin{equation}\label{eq1}
y_{eff}\approx\frac{12\,\left(O/H\right)}{\ln\left(1/\mu\right)}\,,~\textrm{where}~\mu=\frac{\sigma_{\rm{HI}+\rm{H_2}+\rm{He}}}{\sigma_{\rm{HI}+\rm{H_2}+\rm{He}}+\sigma_*}.
\end{equation} 
The radial profiles of the surface densities of gas ($\sigma_{\rm{H_2}+\rm{He}}$ and $\sigma_{\rm{HI}+\rm{He}}$) and stars ($\sigma_*$) were taken from \citet{Leroy2008}. We took into account the bulge contribution to the total stellar surface density profiles. To do it we firstly decomposed the profiles into the components of exponential disk: $\sigma_d(R)=\sigma_{0d}e^{-R/R_d}$ (here $\sigma_{0d}$ is the central surface density of the disk, $R_d$ is the disk exponential scalelength)   and Sersic bulge: $\sigma_b(R)=\sigma_{0b}10^{-b_n(R/R_e)^{1/n}}$ ($\sigma_{0b}$ is the central surface density of bulge, $R_e$ is the effective radius, $b_n$ is defined through the bulge shape parameter $n$). After decomposition we subtracted the bulge contributions from the surface density profiles and used the resulting stellar mass profile in our analysis.

Radial distribution of oxygen abundance:
\begin{equation}\label{eq2}
12+\log\left(\frac{O}{H}\right)=12+\log\left(\frac{O}{H}\right)_0+\rm{C_{O/H}}\cdot\frac{R}{R_{25}},
\end{equation} 
where $R_{25}$ is the optical radius  confined by the B-isophote $25^m$ per arcsec$^2$, was taken from \citet{Moustakas2010}. The term $12+\log\left(\frac{O}{H}\right)_0$ is the intercept and $C_{O/H}$ is the gradient of the relationship. These authors evaluated the oxygen abundances using two strong line calibrations -- the theoretical calibration of \cite{KK2004} (hereafter KK2004) and the empirical calibration of \cite{PT2005} (hereafter PT2005). Note that more recent work by \cite{Pilyugin2014} contains more galaxies with measured O/H than in \citet{Moustakas2010} paper, although the latter authors used a single empirical method for O/H determination without resorting to the emission lines of C or N. 

The difference between theoretical (KK2004) and empirical (PT2005) calibrations cannot be reduced to a constant factor. Although both calibrations have their own weaknesses (see f.e \citealt{Kewley2008}), we  prefer the empirical one. This choice is based on the excellent concordance (within the 0.1dex) between the O/H estimates obtained by the empirical method and by direct ($T_e$) method  for both metal-rich and metal-poor gas (\citealt{Pilyugin2003}), as well as the agreement of gas-phase O/H estimates for the solar circle with the  metallicity of young stars (\citealt{Kudritzki2015}) and with the independent valuation  of O/H from the interstellar absorption line of neutral oxygen (\citealt{Pilyugin2003}).

\begin{figure*}[h!]
\includegraphics[width=6.0cm]{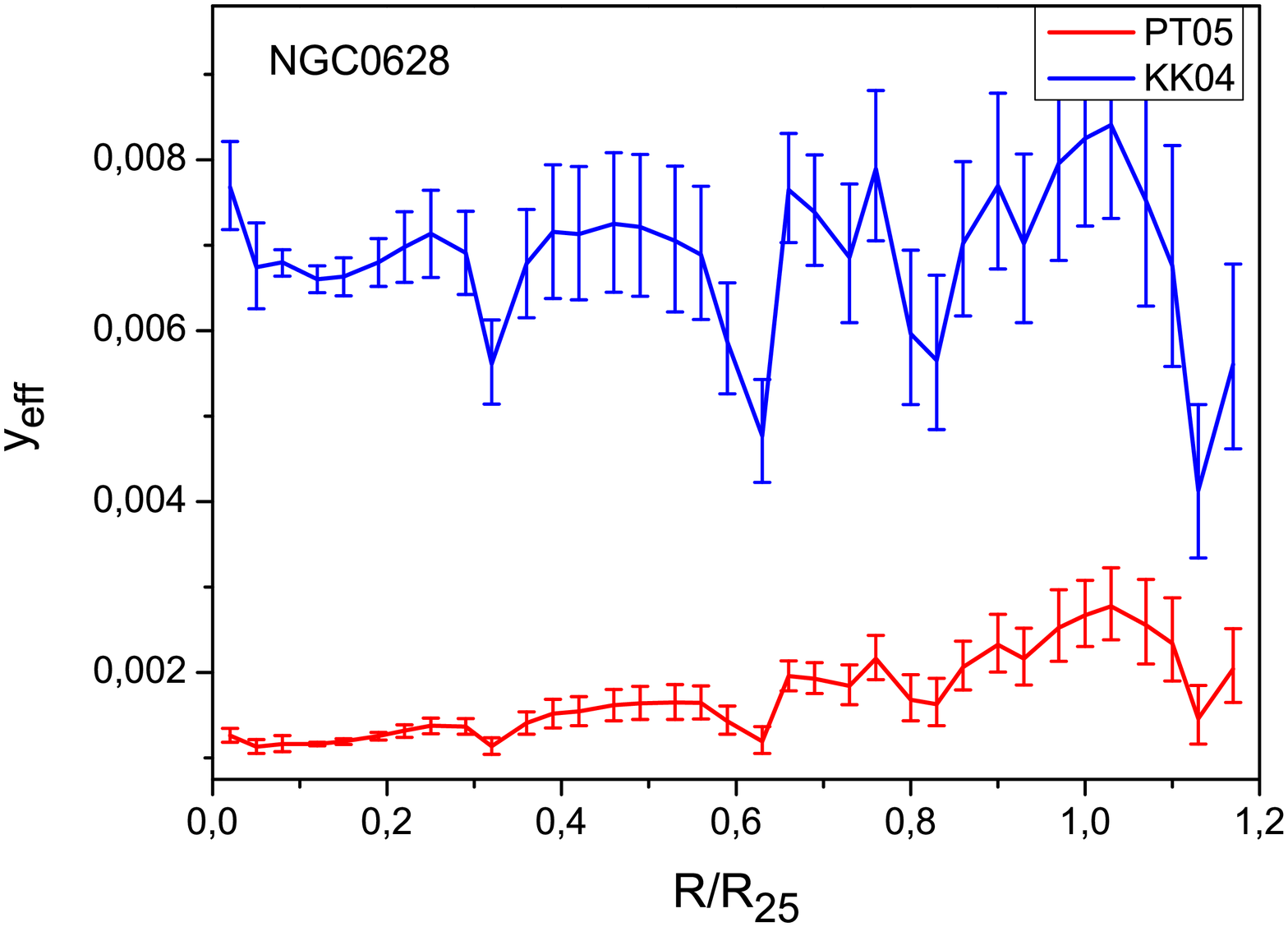}
\includegraphics[width=6.0cm]{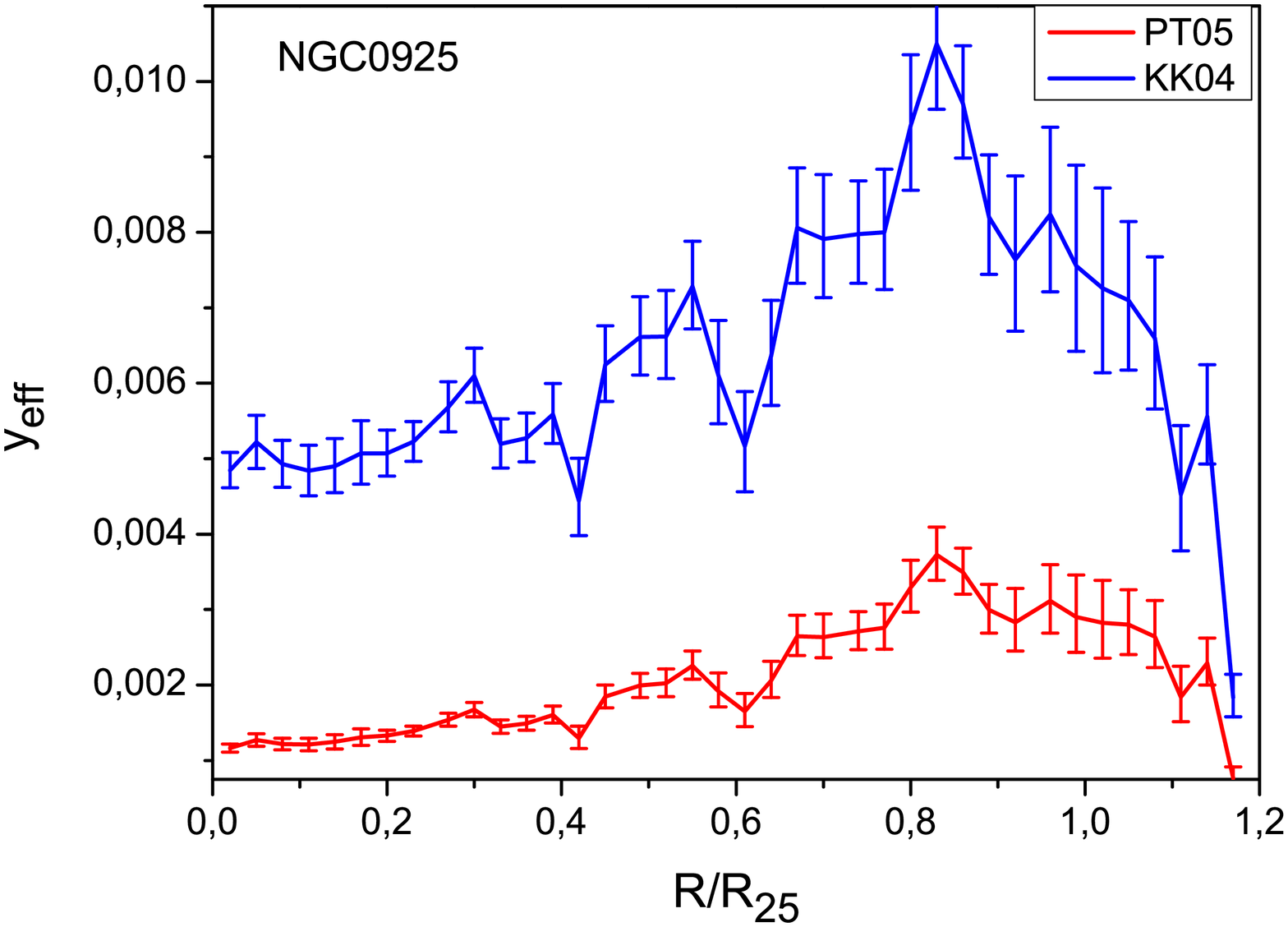}
\includegraphics[width=6.0cm]{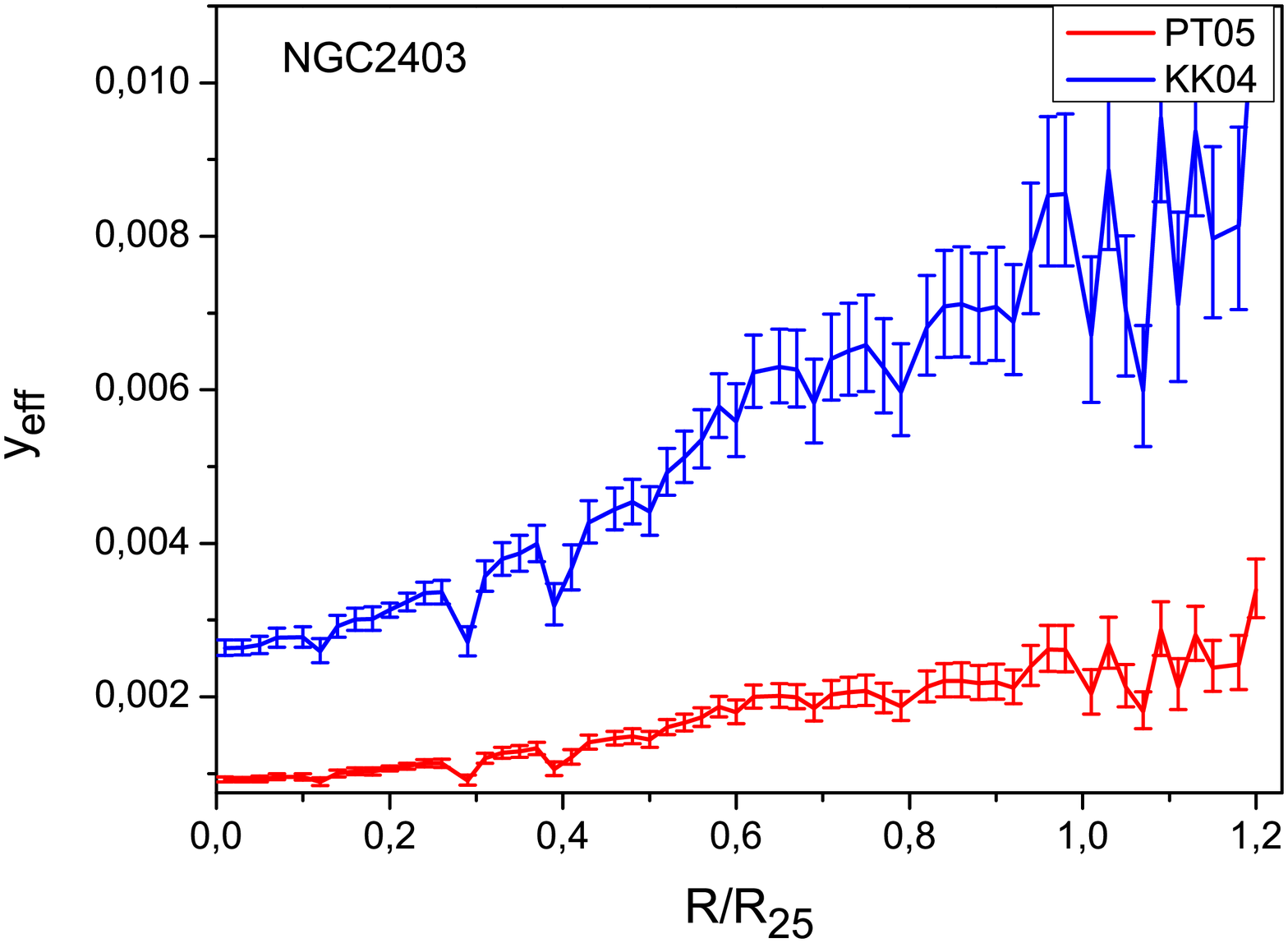}
\includegraphics[width=6.0cm]{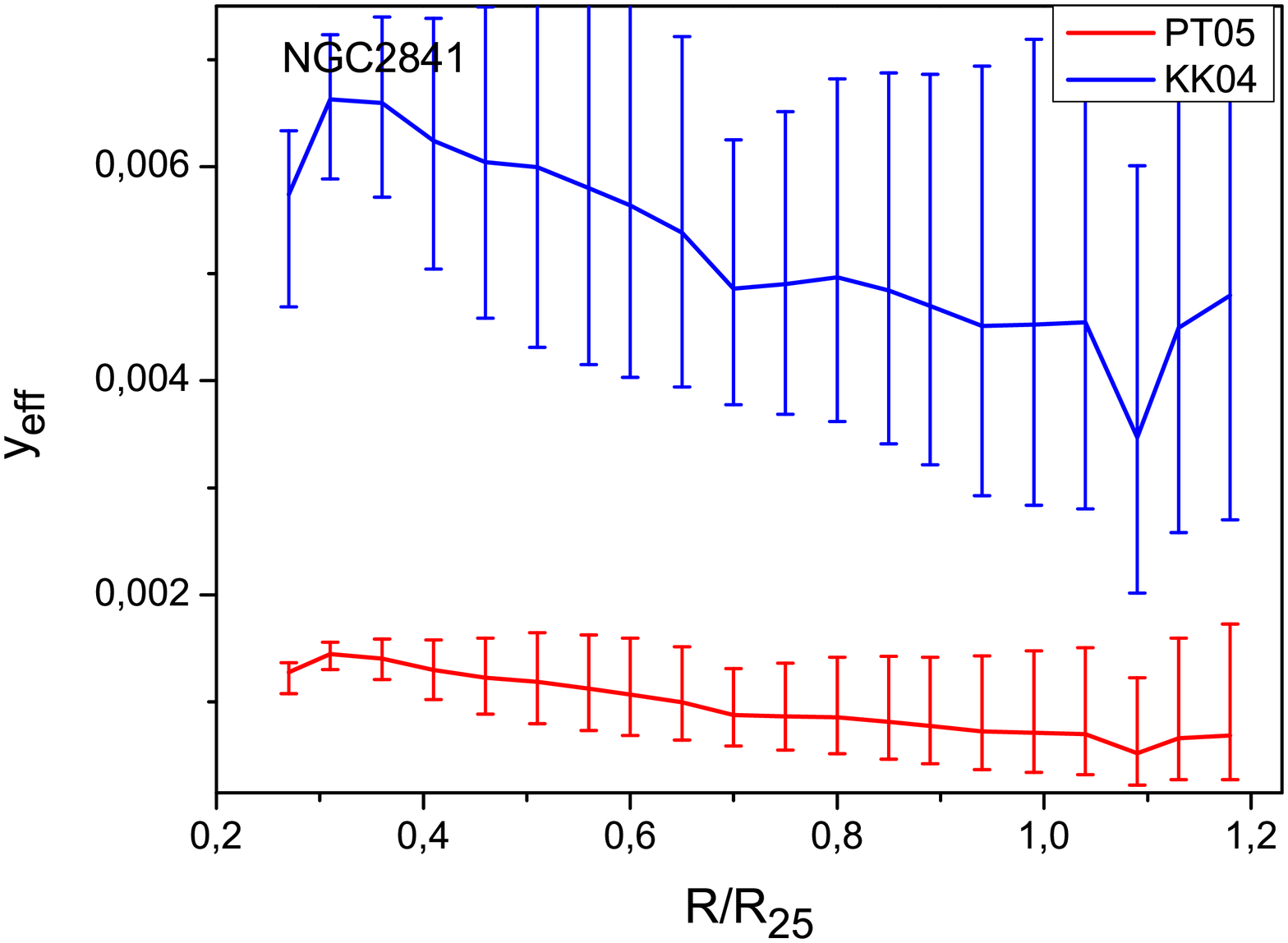}
\includegraphics[width=6.0cm]{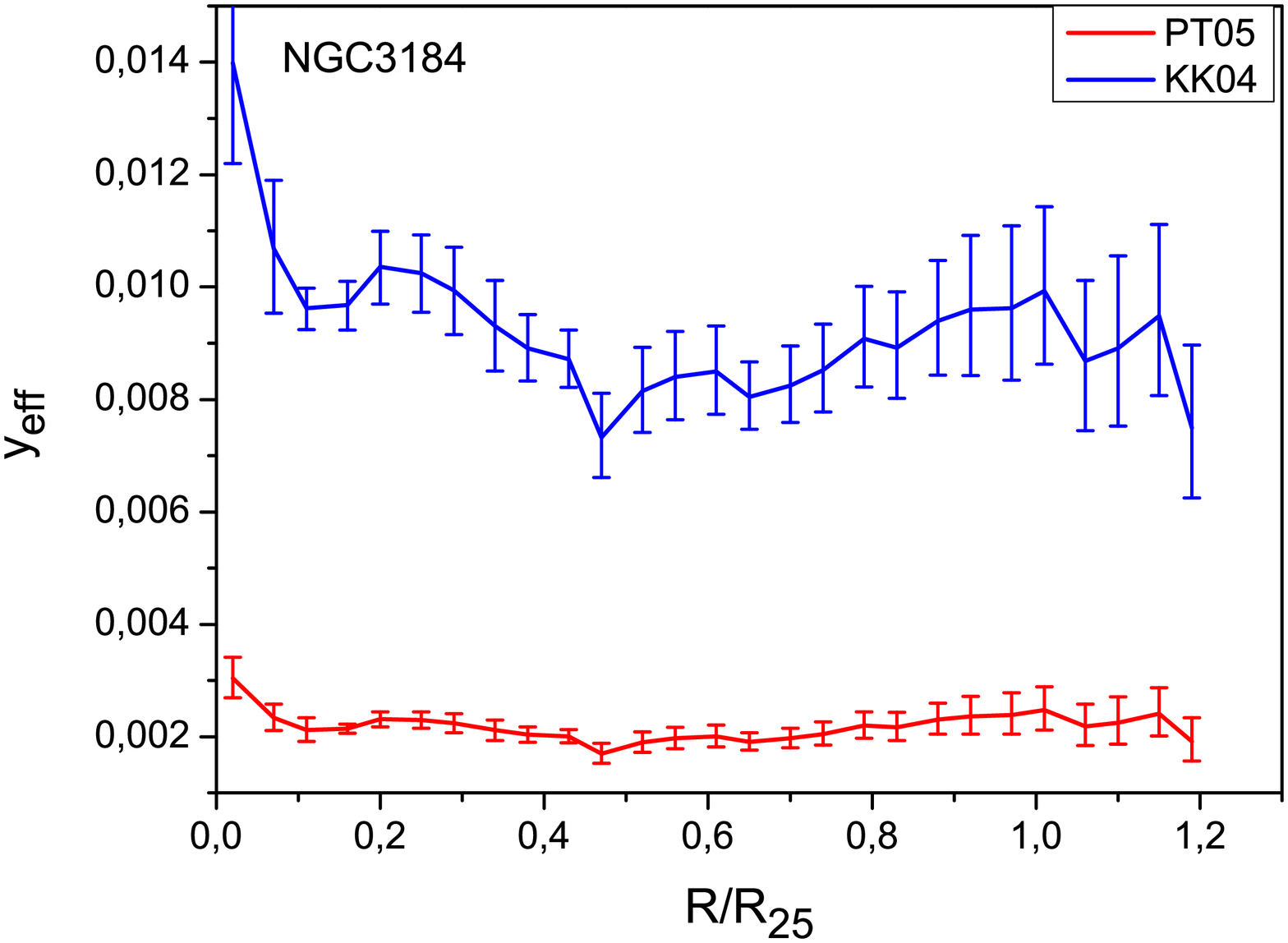}
\includegraphics[width=6.0cm]{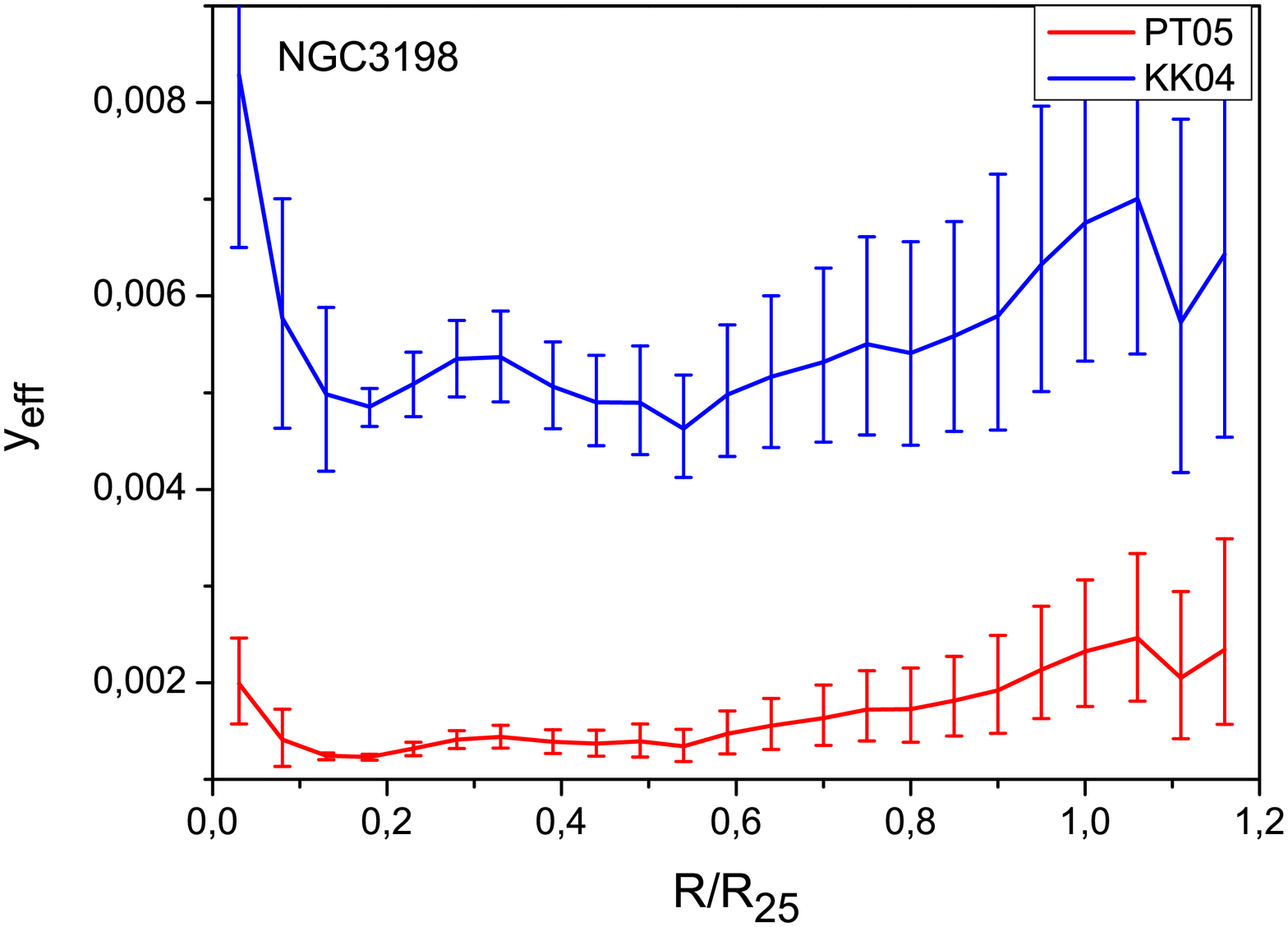}
\includegraphics[width=6.0cm]{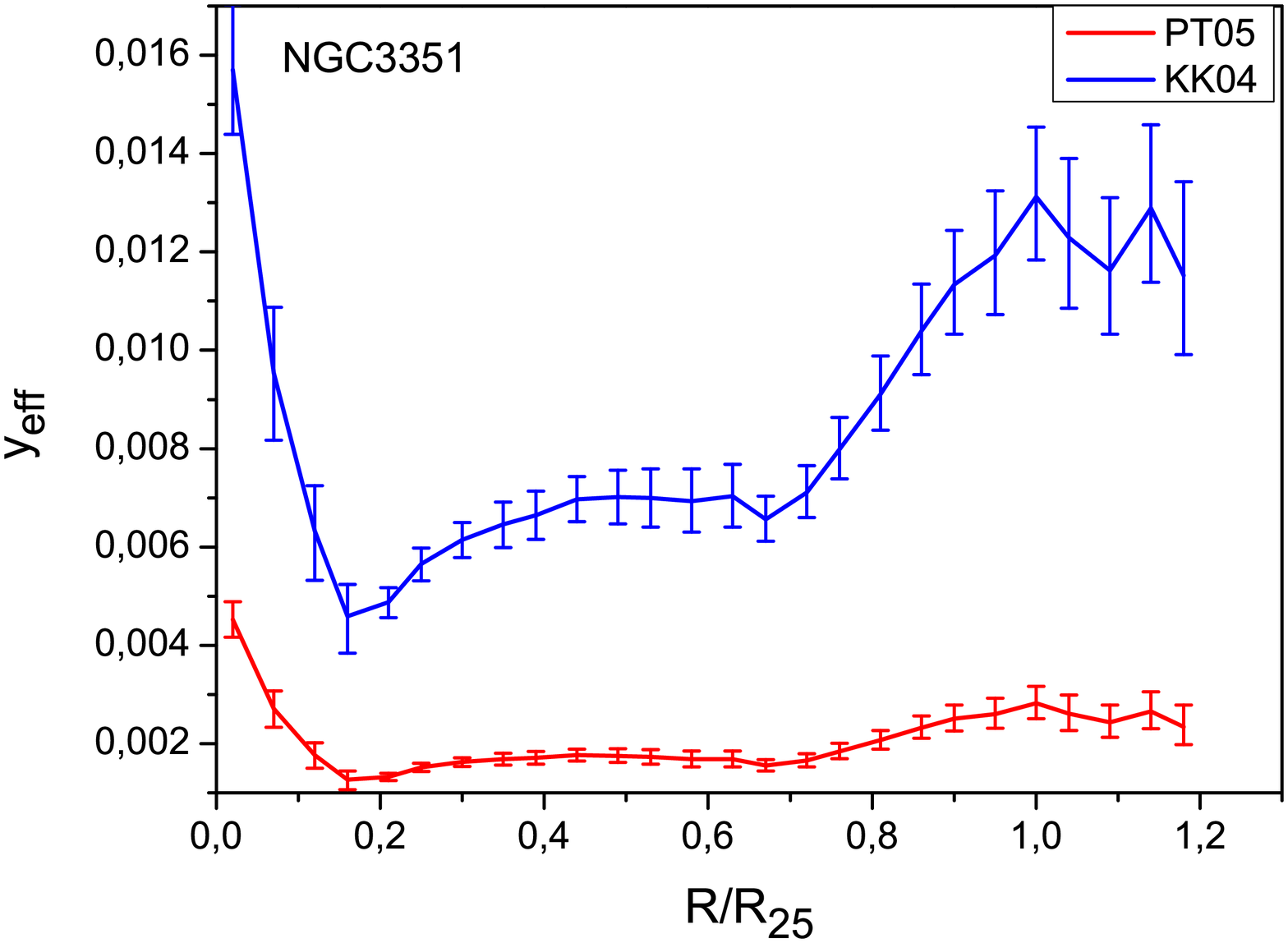}
\includegraphics[width=6.0cm]{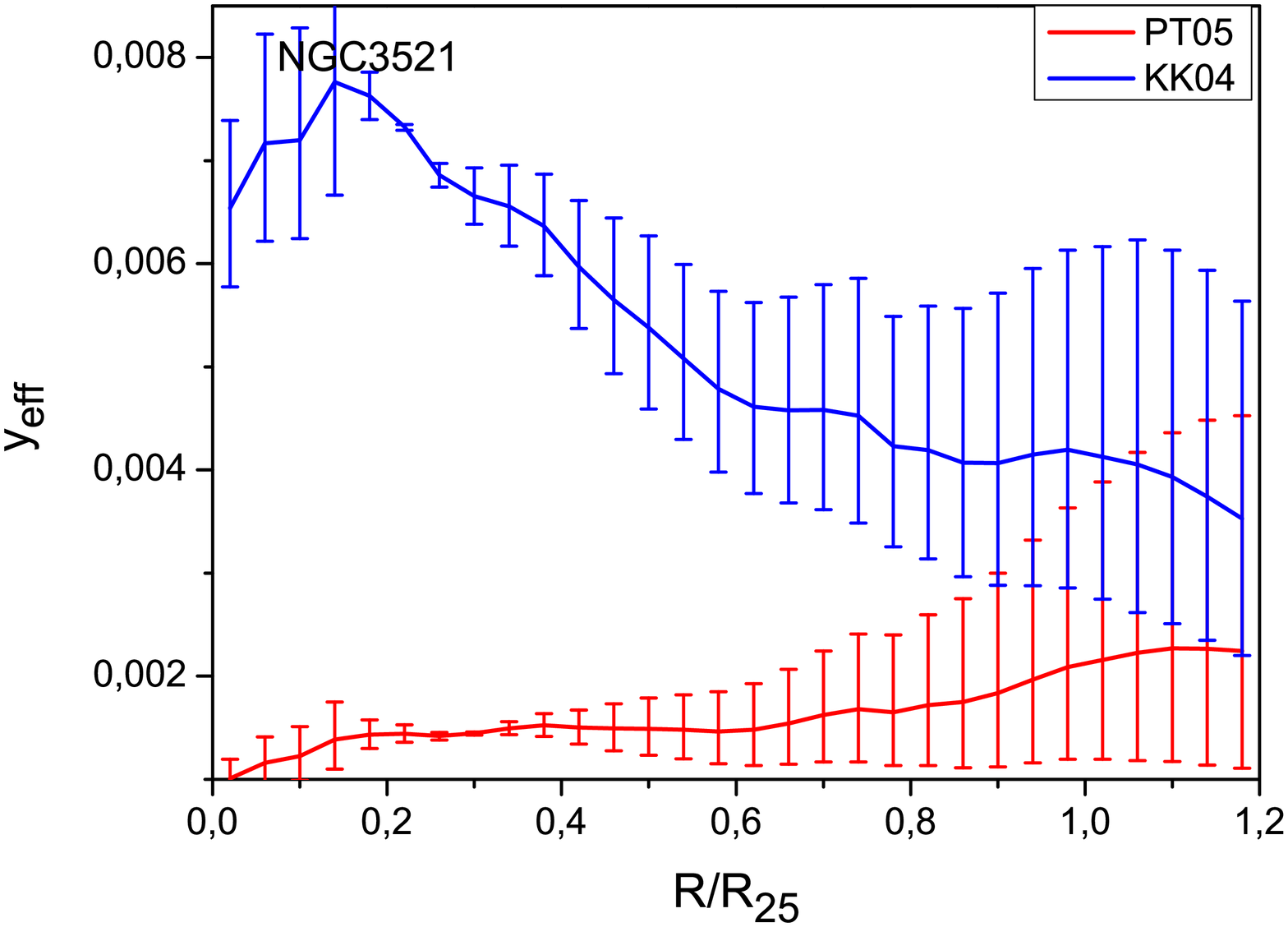}
\includegraphics[width=6.0cm]{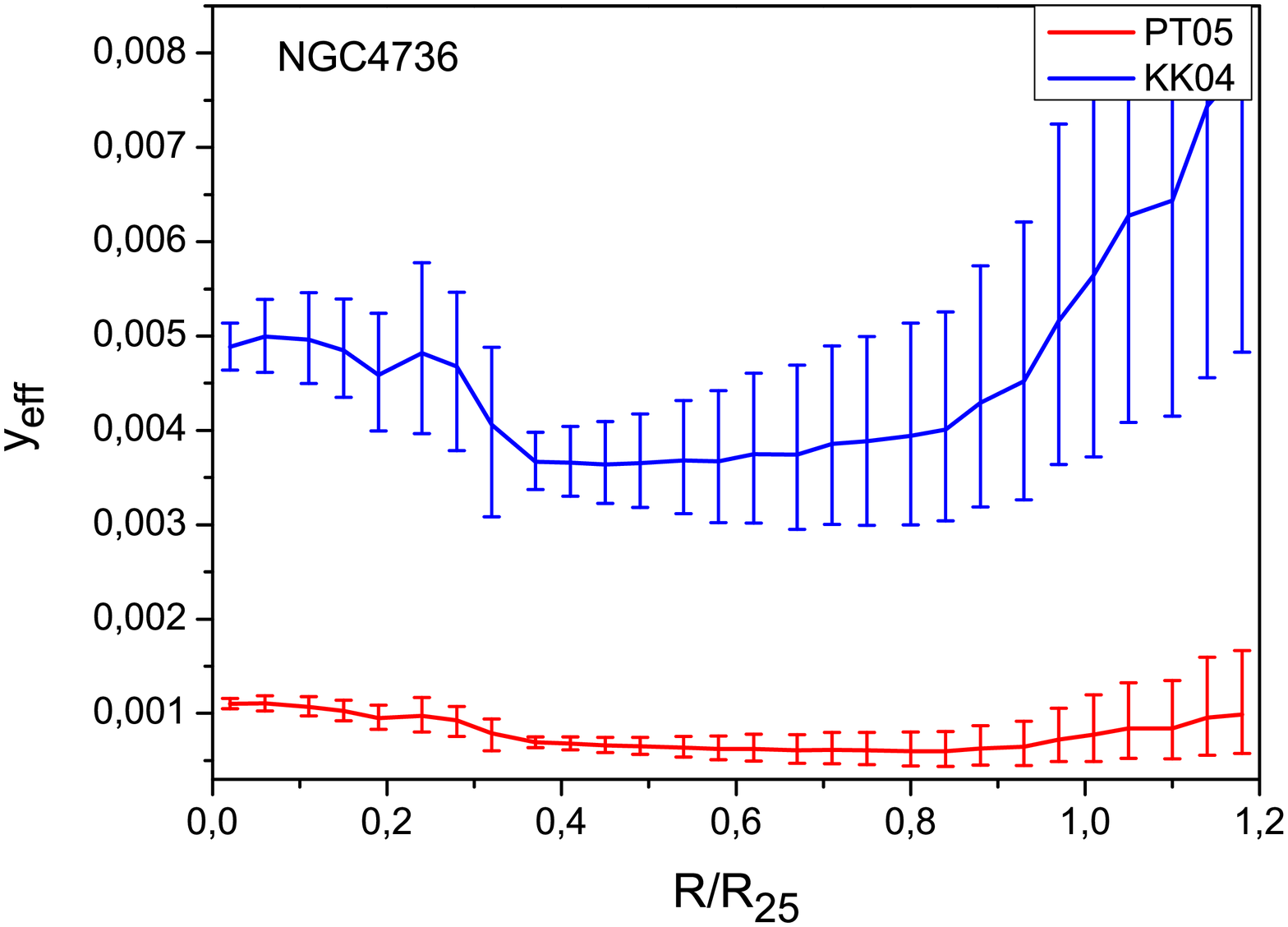}
\includegraphics[width=6.0cm]{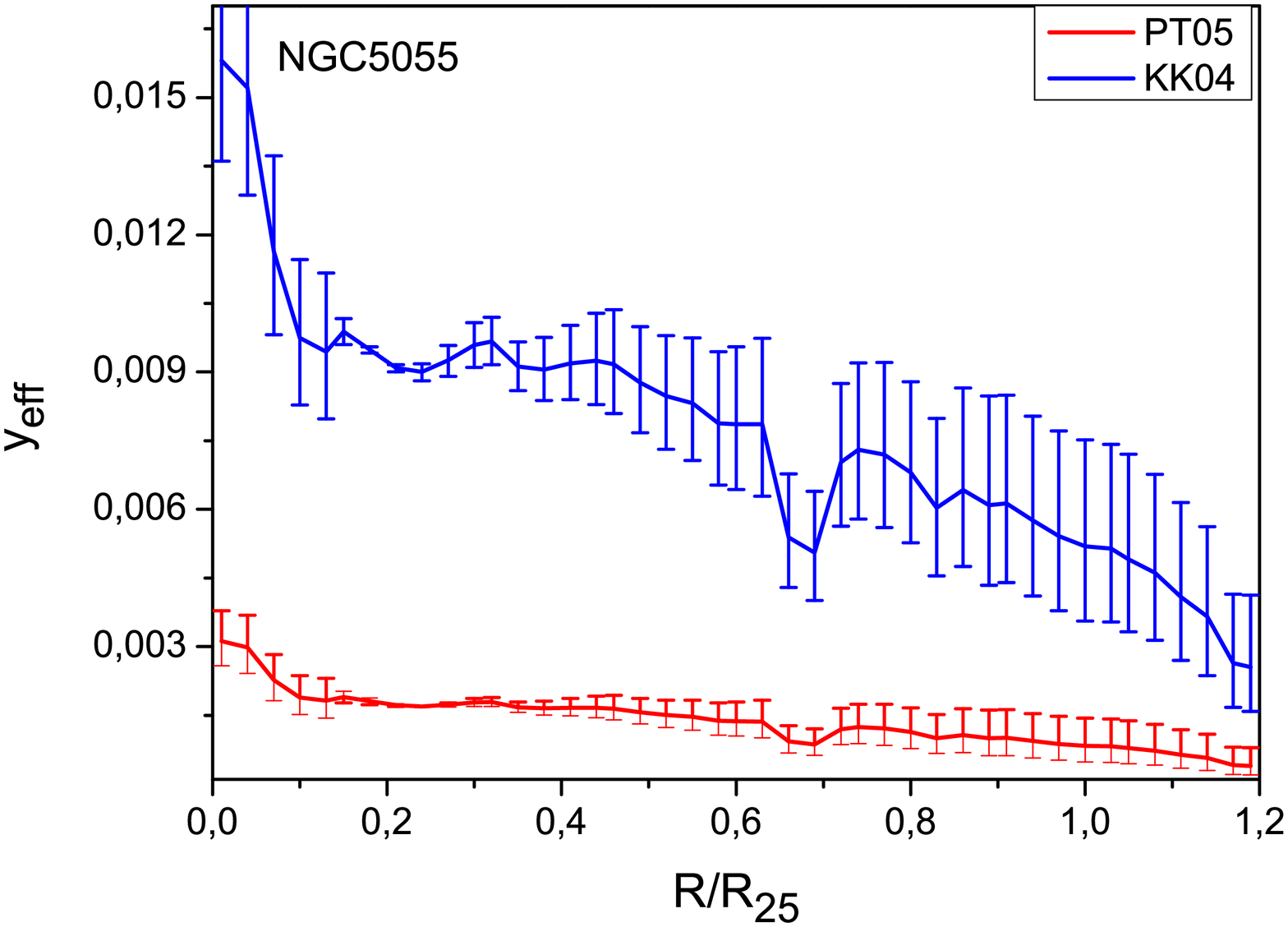}
\includegraphics[width=6.0cm]{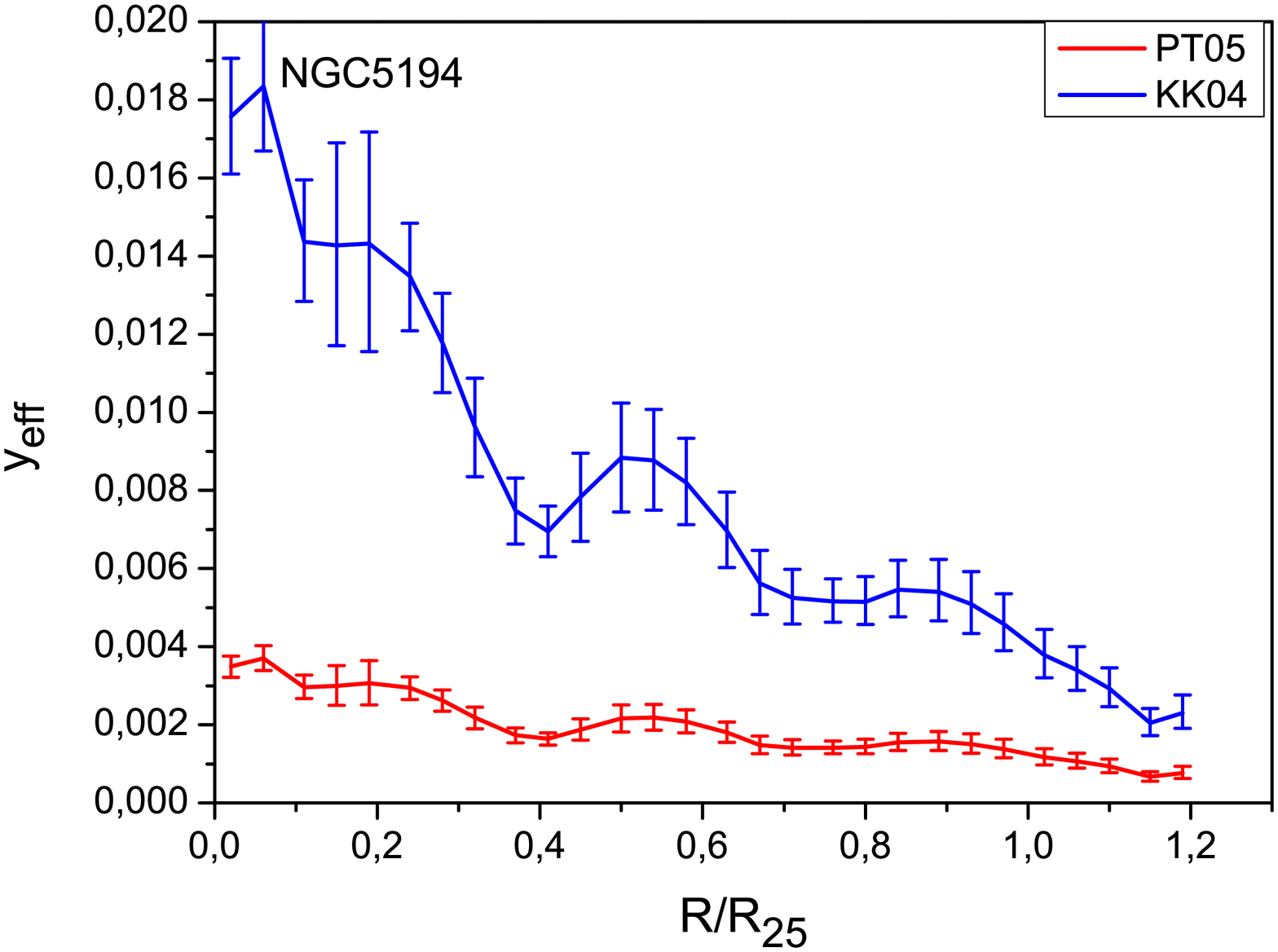}
\includegraphics[width=6.0cm]{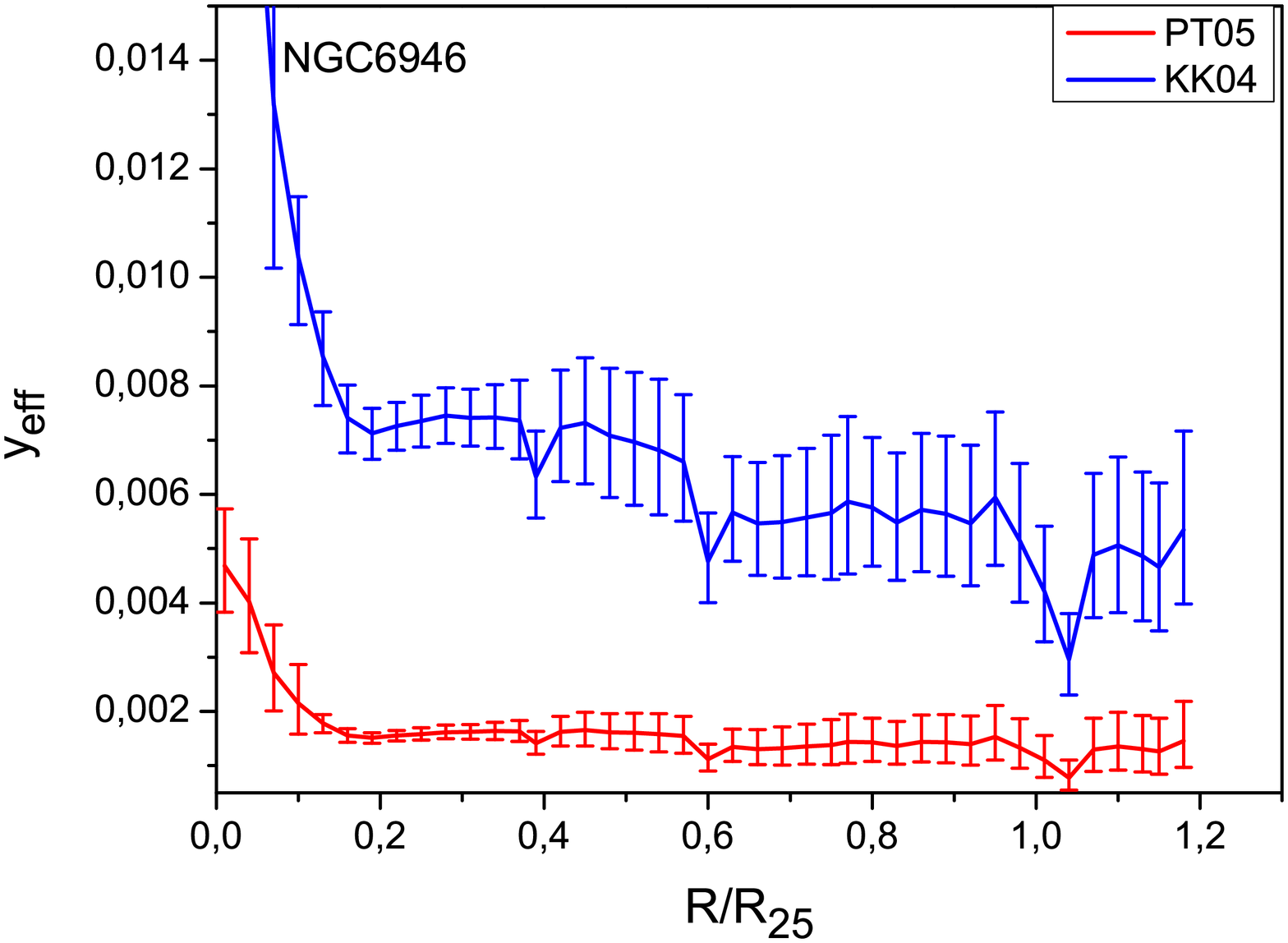}
\includegraphics[width=6.0cm]{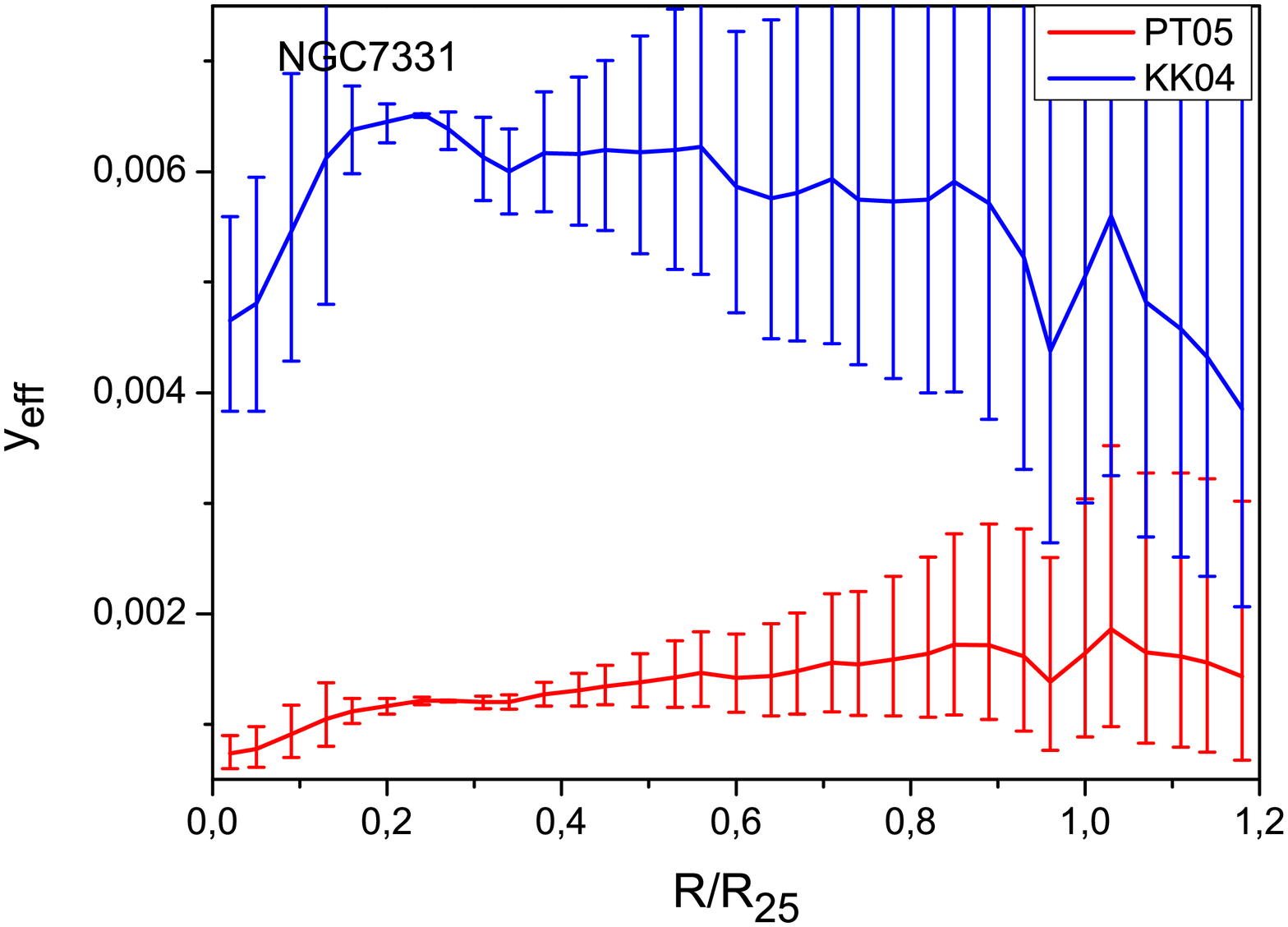}
\includegraphics[width=6.0cm]{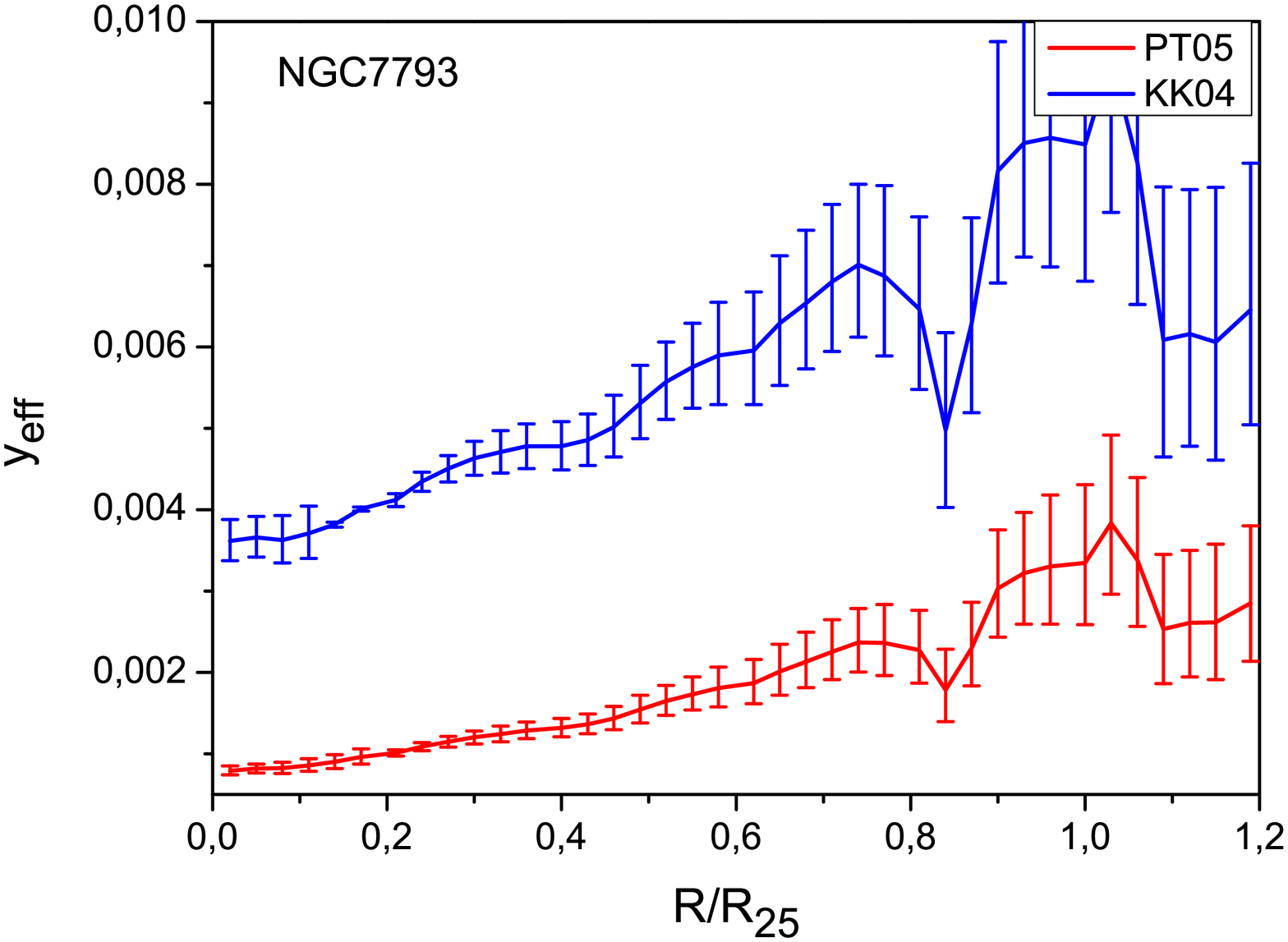}

\caption{Radial profiles of the effective oxygen yield. Top and bottom curves are based on KK2004 and PT2005 methods respectively (see the text).}\label{fig1}
\end{figure*}

 In Fig. \ref{fig1} we show the obtained radial profiles of the effective oxygen yield. As one can expect, the estimations of the effective yield are sensitive to the method of determination of the oxygen abundance. The (O/H) ratios based on the PT2005 calibration  are significantly lower than those obtained by  KK2004 method. The mean ratio of the two yields for our sample galaxies is $4.2\pm 0.96$
and the mean ratio of the gradients is $1.6\pm1.1$.  Despite this systematics, a general shape of radial profiles of $y_{eff}$ has a common behavior in most of galaxies. The ratio of the two yields is roughly constant with radius in a majority of objects from our sample, except NGC3521 and NGC7331 which also have the biggest difference between the oxygen gradients obtained by two calibrations. 

It is essential, that PT2005 calibration gives the effective oxygen yield which increases along the radial distance in most of galaxies of our sample,  although in some cases the growth is very mild so $y_{eff}$ can be interpreted as approximately constant. Taking into account that both gas inflow and outflow reduce  $y_{eff}$, one may conclude that these processes  should be especially effective for the inner parts of most of galaxies considered. As it is shown in Fig. \ref{fig1}, there is a peak of $y_{eff}$ in the innermost region of galaxies,  followed by local minimum of the radial profile. If real, the central maximum may reflect a special way of chemical evolution of the bulge-dominated region, where  a stellar population of  a bulge contributes to the metal enrichment of a gas. Note that the instantaneous enrichment approximation is not valid in this case even in the absence of gas inflow/outflow from a galaxy.

\citet{Belfiore2015} analyzed the profile $y_{eff}$ for NGC0628 -- one of the galaxies from our sample. They used photometry in different bands to obtain a reliable estimate of the stellar mass surface density by
performing full spectral energy distribution (SED) fitting and different metallicity calibrations, and came to qualitatively similar result, that $y_{eff}$ increases to the peripheral disk region.  

There are three galaxies of our sample (NGC 2841, NGC 5055 and NGC 5194) where the effective yield behaves in a different way, being lower for the disk periphery for both PT2005 and KK2004 measurements of $O/H$. KK2004 calibration adds three more galaxies that have a decreasing radial profile of $y_{eff}$ -- NGC3521, NGC6946 and NGC7331, demonstrating that the radial behavior of $y_{eff}$ depends on the calibration used. All of these 6 galaxies have high total luminosity and velocity of rotation. We did not find any connection between the presence of a bar or a ring or the galaxy environment and the shape of the radial profile of $y_{eff}$\footnote {It should be noted, however that one of these galaxies - NGC5194 -  is a strongly interacting galaxy, which may have an effect on the chemical evolution of its disk.}. These galaxies do not stand out also by their dark halo masses inside of optical borders presented in \citet{SaburovaDelPopolo}: their dark halo to stellar mass ratios are quite typical being in a range 0.9-2.4.
It is worth noting that the mean central metallicity  of  three galaxies with the decreasing of effective yield for both PT2005 and KK2004 calibrations is $12+\log(O/H)_0\approx 8.50$  (PT2005) and $\approx 9.2$ (KK2004), which is slightly higher than for the rest of galaxies of our sample: $12+\log(O/H)_0\approx 8.37\pm 0.1$ (PT2005) and $12+\log(O/H)_0\approx 8.98\pm 0.13$ (KK2004). It is natural to propose that the mechanisms reducing $y_{eff}$ along the radius  in these galaxies, unlike the other ones,  more effectively influence the gas-phase metallicity at large radial distances than in the central parts, reflecting some peculiarities of their evolution (say, due to minor merging, accretion or/and outflow of enriched gas).

 As far as the highest
observed values of $y_{eff}$  should be restricted by the level of true nucleosynthetic yield $y_o$, the saturation level of   $y_{eff}$  for different galaxies may be used  to estimate $y_{o}$ empirically. By this method  it was found $y_{o}$ = 0.003 - 0.004 (\citealt{Pilyugin2007}; \citealt{Dalcanton2007}). Similar value was obtained for the Milky Way disk  by \cite{Kudritzki2015} in their analytical chemical evolution model with a constant ratios of galactic mass-loss and mass-gain
to the star formation rate. Taken as the upper limit, its value agrees pretty well with $y_{eff}$ found for the galaxies considered here by PT2005 method of oxygen abundance estimate, however it is much lower than the values followed from KK2004 estimates (see Fig. \ref{fig1}). It is essential that irrespectively on where $y_{eff} $ is maximal - in the central parts or at large radial distances - its values  based on PT2005 calibration nowhere exceed the estimates of  $y_o$ mentioned above.

It is worth noting that the recent paper by \cite{Vincenzo2015} 
which considers stellar evolution models for not-too-low metallicity $Z / Z_\odot \ge 10^{-3}$ gave much higher values $y_{o}$= 0.01-0.04 depending on the adopted stellar initial mass function (IMF). The difference between these large values of $y_o$ and the estimates of $y_{eff}$ for PT2005 data is too big to be compatible with each other: either the  true yield theoretically found by \cite{Vincenzo2015} is strongly overestimated, or the semi-empirical method of measurements of the oxygen abundance strongly underestimates (O/H) and $y_{eff}$. One should have in mind however, that the stellar evolution- based estimates  of  $y_o$ still remain not too reliable. Note also that even the KK2004- values of $y_{eff}$ are in most cases lower than 0.01.

There are not so many known mechanisms which may reduce significantly $y_{eff}$ in comparison with the expected stellar yield $y_o$. These are:\\
i) low upper mass limit of IMF;\\
ii)  bottom-heavy stellar IMF; \\	
iii) accretion  onto a disk of low abundant gas from the intergalactic medium (cold flows) or from a halo (hot accretion mode of cooling gas or the recycled galactic wind; see \citealt{Oppenheimer2010}), or a gain of non-evolved gas as the result of minor merging;\\
iv) outflows of the enriched gas induced by star formation from a disk to halo or into the intergalactic space.

The assumption of the lower upper limit of mass of forming stars  or the use of the IMF with a shallower slope in the high mass end will certainly reduce $y_{eff}$ (see the numerical estimates by \citealt{Vincenzo2015}). As it was demonstrated by \cite{2014arXiv14046533M} the IMF  may vary along the galactocentric distance, however this question requires further investigation. If, on the other hand, to assume that IMF is more bottom-heavy than it is usually accepted, it will not change $y_{eff}$	 significantly, however it should decrease the intrinsic stellar yield $y_o$ as the result of decreasing of the mass ratio of  massive-to-low-massive stars. In this case one can expect that the mass-to-light ratios of stellar population of disks will reveal a tendency to be lower in the inner regions, where  $y_{eff}$ is usually the most low. However this version seems to be unlikely, being in conflict with the estimations of mass and mass-to-light ratio of stellar disks at different radii obtained by the decomposition of rotation curves (\citealt{Martinsson2013}). 
 
 Outflow of gas enriched  by supernovae and the accretion of  non-enriched gas give similar effect on the metal abundance, wherein their roles may be different in the inner and in the outer disk regions. Gas outflow  reduces the  radial O/H distribution more effectively than the accretion if the relative gas content $\mu$ is close to unit (\citealt{Dalcanton2007}), and hence it may be most essential in the peripheral gas-rich regions, where the relative mass of a gas is higher.

Accretion evidently plays a significant role in the chemical evolution of galaxies, but it hardly can explain the $y_{eff}$ which is several times lower than the true yield if the gas fall is just a recent episode: in this case, even under most favorable condition, when the accreted non-enriched gas does not participate in star formation, the  total amount of the accreted gas should several times exceed the initial mass of gas (\citealt{Dalcanton2007}).  In a more realistic case, when the accretion is the long-lasting process, with the rate, say, proportional to star formation rate,  the effect of accretion may be significant, especially if the gas to stellar disk mass ratio $\mu$ is low. The model developed by \cite{Kudritzki2015} shows that for  the  accretion rate twice as low as SFR and  for $\mu$ = 0.02 - 0.4, which is typical for our galaxies,  the abundance of oxygen should be 2 - 4 times lower than that expected  for closed box model (see Fig.1 in \citealt{Kudritzki2015}). Indeed, $\mu$ usually steadily grows to the disk periphery of galaxies because the gas (mostly H I) surface density decreases along the radius more slowly than the stellar disk surface density. Hence, the accretion rather than the outflow may be responsible for  the observed fall of $y_{eff}$ in the central regions of galaxies, where the gas mass fraction $\mu$ is relatively low.  In this case the metal abundance is expected to be close to that for closed box model (\citealt{Kudritzki2015}). This result is not sensitive to the choice of model parameters of the evolution (see also \citealt{Ascasibar2015}).  

The other way the gas infall may reduce a metal abundance and hence $y_{eff}$ in the inner disk regions is the hot mode accretion of the cooling non-enriched halo gas, induced by the galactic fountains which inject a gas to halo as a feedback of star formation (see \citealt{Fraternali2006}). Indeed, star formation rate is  usually the most intense in the inner galaxy, hence one may expect that the accretion may be more essential there.

A decreasing of gas-phase metallicity in the central disk regions with respect to what is expected for the closed box model  should lead to the  flattening of the abundance gradient.  A reduction of metallicity due to accretion may also be responsible for the local decline of grad (O/H) in the central region, directly  measured  in some galaxies (\citealt{Belfiore2015}; \citealt{Sanchez2014}).

\begin{table}[h]

\caption{Galaxy sample.} \label{tab}
\begin{tabular}{@{}llll@{}}
\hline%
Galaxy&Distance, Mpc&Type&$M_{B}$\\
\hline%

  NGC0628        	&	7.3	&	 Sc              	&	-20	\\
  NGC0925        	&	9.2	&	 Scd             	&	-20	\\
  NGC2403        	&	3.2	&	 SABc            	&	-19.4	\\
  NGC2841        	&	14.1	&	 Sb              	&	-21.2	\\
  NGC3184        	&	11.1	&	 SABc            	&	-19.9	\\
  NGC3198        	&	13.8	&	 Sc              	&	-20.8	\\
  NGC3351        	&	10.1	&	 Sb              	&	-19.9	\\
  NGC3521        	&	10.7	&	 SABb            	&	-20.9	\\
  NGC4736        	&	4.7	&	 Sab             	&	-19.8	\\
  NGC5055        	&	10.1	&	 Sbc             	&	-21.1	\\
  NGC5194        	&	8	&	 Sbc             	&	-21.2	\\
  NGC6946        	&	5.9	&	 SABc            	&	-20.6	\\
  NGC7331        	&	14.7	&	 Sbc             	&	-21.7	\\
  NGC7793        	&	3.9	&	 Scd             	&	-18.8	\\
\hline%
\end{tabular}
\end{table}

\section{On the relation between maximal oxygen yield, dark halo and total masses within the optical radius}

A dark matter halo plays the important role in the evolution of the disks affecting both the accretion rate and gas losses and hence influencing the gas-phase metal abundance. Numerical simulations show that for the low redshifts the mechanism of accretion is different for galaxies with massive and low massive dark halos (see discussion in \citealt{Bouche2010}; \citealt{Oppenheimer2010}; \citealt{Almeida2014}).  For galaxies with massive dark halo a falling gas is shocked and virialized, so that the accretion comes from  hot halo gas after its cooling or from material that was previously thrown out of a galaxy into a halo. For massive halos the temperature of halo gas is higher and gas condensation is regulated by thermal cooling rate, so it needs billions years for gas to cool, while for less massive galaxies cooling flows  of partially ionized gas  may directly reach the disk, being the principal mode of accretion. It does not mean that massive halos are guaranteed to provide a low accretion rate, because they   may contain a huge mass of gas with the cooling time comparable or lower than the age of galaxies, especially if to take into account the galactic fountains inspiring a gas cooling (see \citealt{Fraternali2006}).

It is worth comparing the observed $y_{eff}$ of galaxies with the masses of dark halo for concrete galaxies.  Note that there are two definitions of a halo mass which are in use. A total, or virial dark halo mass is the mass within the virial radius which is  well outside of the optical radius.  This mass comes from cosmological scenarios of formation of galaxies and is tightly linked with the circular velocity at the virial radius, which does not differ much from the maximal velocity of rotation $V $ of a disk, being slightly lower than the latter (see f.e. \citealt{Reyes2012}).  Unlike the virial mass, a directly measured halo mass is usually restricted by the radius of the observed disk. To be specific, it is convenient to define a halo mass inside of $R_{25}$ which is the order of magnitude lower than the virial mass of halo.

 It is well known that the gas abundance is higher in massive and fast rotating galaxies that is in those with massive halos. However this trend may be a result of different history of star formation, so its connection with $y_{eff}$ is not evident.  \cite{Dalcanton2007} found a positive correlation between   $V $ and $y_{eff}$ at  R=0.4$R_{25}$, using the data from \cite{Pilyugin2004}, however this correlation is clearly seen only for dwarf galaxies with $V <100$ $km ~s^{-1}$, where the effects of gas outflow may be essential (see however \citealt{Gavilan2013}, which found yield $y_{eff} \sim 0.006$ with high dispersion  for gas-rich dIrr galaxies). The galaxies we consider here rotate much faster. We do not find for them any significant link  between $V $ and $y_{eff}$ taken neither at R = 0 nor at R = 0.4$R_{25}$ (a correlation coefficients are less than 0.5).

\begin{figure*}[t]
\includegraphics[width=9.0cm]{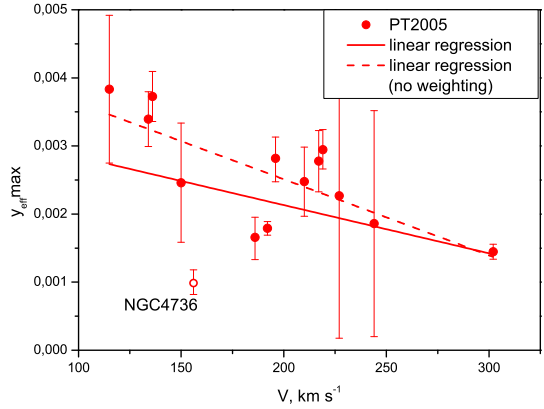}
\caption{The maximal effective yield for $O/H$ estimates based on the calibration of PT2005 compared to the rotation velocity taken from \citet{Leroy2008}.   The dashed and solid lines show the linear regression with and without taking into account the error bars respectively (NGC4736 was ignored).  }\label{vyeff}
\end{figure*}

Curiously, a correlation appears for fast-rotating galaxies if we take $y_{eff,max}$ instead of $y_{eff}(0)$, where $y_{eff,max}$ is the maximal value of $y_{eff}$ for a given  galaxy. (We did not take into account the central peak of $y_{eff}$ in the bulge-dominated regions at $R<0.2R_{25}$ where the accretion regime is dictated by bulge contribution). In other words, $y_{eff,max}$  is the effective yield at such radial distance of the main disk where one can expect the weakest influence of accretion and/or gas outflow on the gas-phase metallicity. In most, but not in all cases, $y_{eff}$ reaches its maximum in the outer disk (some exceptions were discussed in Section 2). Fig. \ref{vyeff} demonstrates that for spiral galaxies considered here, where, unlike to dwarf galaxies, the effect of gas outflow is expected to be relatively low, there is a tendency to have lower values of $y_{eff,max}$ for the most rapidly rotating galaxies, possessing massive virial dark halos. The numerical information for all correlations is given in Table \ref{tab2}. In the chi-square reducing procedure we assigned the statistical weights  $w_i=1/s_i^2$ where $s_i$ is the error bar of a given point. We performed the reducing procedure both with and without weighting (solid and dash lines correspondingly).   
 As far as a massive halo tends to reduce galactic winds, it allows to conclude  that it is the accretion which plays a major role in the more massive fast rotating galaxies containing a large reservoir of slowly cooling gas.  
 
 In this and other diagrams for $y_{eff, max}$ values, we used the oxygen abundance measurements by PT2005. The results for  KK2004 calibration, which we do not reproduce here, give qualitatively similar relationships, although with larger spread of points on the diagrams.  

\begin{figure*}[t]
\includegraphics[width=9.0cm]{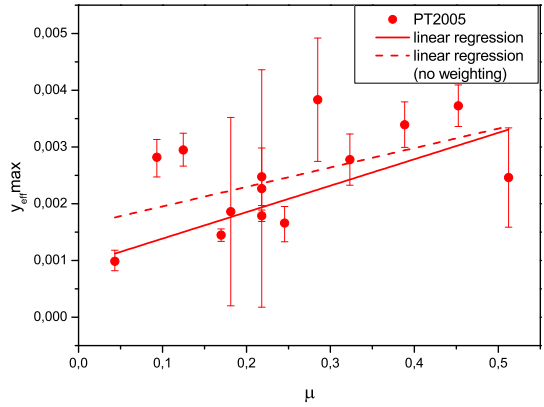}
\caption{The maximal effective yield for $O/H$ estimates based on the calibration of PT2005 compared to the total gas mass fraction taken from \citet{Leroy2008}. The dashed and solid lines show the linear regression with and without taking into account the error bars respectively. }\label{muyeff}
\end{figure*}

\cite{Dalcanton2007} found the anti-correlation between the gas mass fraction and $y_{eff}$, however it is evident only after the inclusion of dwarf irregular galaxies. For the spiral galaxies we consider here such correlation is absent. Instead we found a positive trend (although with high point spread) between the maximal effective yield and the total gas mass fraction  $\mu$ (see Fig. \ref{muyeff}). This relation indicates that the deviation of the effective yield from the real stellar yield is stronger for galaxies with lower gas content: in this case a chemical evolution of gas is more sensitive to the accreting gas flow.
    
\begin{figure*}[h!]
\includegraphics[width=9.0cm]{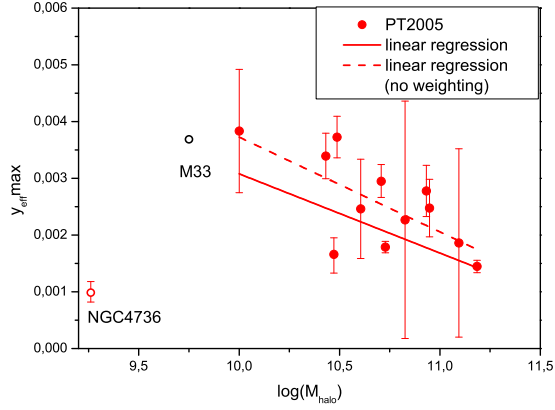}
\caption{The maximal effective yield for $O/H$ based on the calibration of PT2005 compared to the dark halo mass within the optical radius. The dashed and solid lines show the linear regression with and without taking into account the error bars respectively  (galaxies marked by open symbols are ignored). Open circle shows the position of M33 \citep{SaburovaZasov}.}\label{mhyeff}
\end{figure*}

Since the behavior  of $y_{eff}$ beyond the optical borders is not known, 
it is worth trying to compare $y_{eff,max}$ with the dark halo mass $M_{halo}$ inside of optical radius $R_{25}$, using the estimates of $M_{halo}$ compiled in \citet{SaburovaDelPopolo}. The results are given in Fig. \ref{mhyeff}. They demonstrate the anti-correlation between the halo mass and $y_{eff,max}$, which is in agreement with the proposed growing role of accretion along the halo mass sequence. Black open circle in Fig. \ref{mhyeff} marks the position of M33 according to \citet{SaburovaZasov} for $O/H$ ratio determined using $T_e$-method. It agrees with general tendency to have a higher $y_{eff,max}$ for less massive spiral galaxies.

\begin{figure*}[h!]
\includegraphics[width=9.0cm]{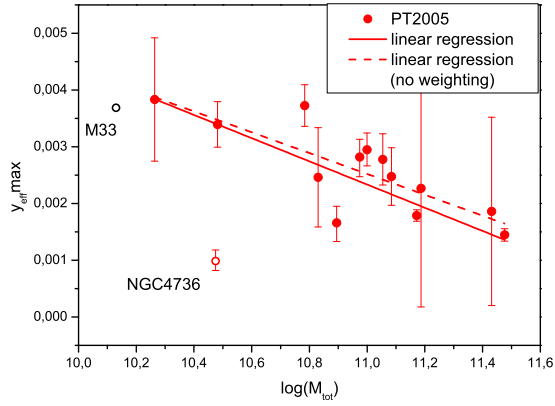}
\caption{ The maximal effective yield for $O/H$ estimates based on the calibration of PT2005 compared to the total mass within the optical radius. Black circle shows the position of M33 \citep{SaburovaZasov}.  The dashed and solid lines show the linear regression with and without taking into account the error bars respectively (galaxies marked by open symbols are ignored).}\label{mtotyeff}
\end{figure*}

 The correlation with  $y_{eff,max}$ becomes tighter  if instead of a dark halo mass we take the total mass of a galaxy within $R_{25}$: $M_{tot}=R_{25}V^2/G$ (see Fig. \ref{mtotyeff}). The outlier in the diagrams is NGC4736. This galaxy possesses the lowest mass of dark halo among all the galaxies we consider (its rotation curve falls to the disk periphery) and hence it has a shallow potential well, so that the effective gas outflow could be responsible for the low value of the effective yield.
A position of this galaxy on the diagram agrees with that of low-mass dIrr galaxies, which occupy a very wide range of $y_{eff}$ revealing a positive correlation with the velocity of rotation (\citealt{Dalcanton2007}), as expected in leaky box models (see however the discussion in \citealt{Gavilan2013}). It seems that the spiral galaxies have a different mechanism regulating the gas inflow/outflow than the dwarf ones.

We came to conclusion that the maximal values of the effective yield are systemically lower for more massive dark halos or more massive spiral galaxies. Low maximal value of the effective yield  $y_{eff,max}$ means that the gas-phase abundance in a given galaxy  is  significantly reduced in comparison with the closed box model all-over a disk.  As far as the gas outflow is less effective for more massive galaxies, it is natural to propose that  the accretion of metal-poor gas, apparently via a cooling of hot halo gas,  plays the most important role in galaxies which possess a high total or dark halo mass. It agrees with the models of formation and a subsequent evolution of galaxies, which predict a growth of the efficiency of accretion with a halo mass.  It is worth mentioning the model of formation of disk galaxies in the preheated media, developed by \cite{Lu2015}, which enables to reproduce remarkably well a number of observational scaling relations. In this model the intergalactic circum-halo gas is assumed to be preheated up to a certain entropy level before it is accreted into dark matter haloes (the preventative feedback model). Such approach strongly reduces the baryon fraction that can collapse into the low mass halos. In agreement with our data, the model predicts a monotonous increasing of the gas accretion efficiency along the dark halo mass sequence for the present day galaxies (see Fig.10 in \citealt{Lu2015}).


\begin{table*}[h!]
\caption{The linear regression equations. (1) -- the linear regression equation; (2) -- the interception ; (3) -- the slope; (4) --  the correlation coefficient $r$; (5) -- note on the taking into account of the error bars (yes for usage of weighting in the estimation of the regression, no for no weighting used). \label{tab2}}

\begin{tabular}{llllll}
    \hline
Equation&A($\cdot 10^{-3}$)&B ($\cdot 10^{-3}$)&$r$&weighting\\
\hline
(1)&(2)&(3)&(4)&(5)\\
\hline
$ y_{eff~max}=A+B\log(M_{halo}(R_{25})) $&$17\pm6$&$-1.4\pm 0.5$&0.62&yes\\
$ y_{eff~max}=A+B\log(M_{halo}(R_{25})) $&$20\pm6.0$&$-1.7\pm 0.6$&0.68&no\\

\hline
$ y_{eff~max}=A+B\log(M_{tot}(R_{25}))$&$25\pm4.8$&$-2.0\pm 0.4$&0.82&yes\\
$ y_{eff~max}=A+B\log(M_{tot}(R_{25}))$&$23\pm4.95$&$-1.8\pm 0.4$&0.80&no\\

\hline
$ y_{eff~max}=A+BV$&$3.5\pm0.6$&$-0.007\pm 0.002$&0.65&yes\\
$ y_{eff~max}=A+BV$&$4.7\pm0.6$&$-0.01\pm 0.003$&0.74&no\\

\hline
$ y_{eff~max} =A+B\mu $&$0.9\pm0.4$&$4.6\pm 1.7$&0.61&yes\\
$ y_{eff~max} =A+B\mu $&$1.6\pm0.4$&$3.4\pm 1.6$&0.53&no\\

\hline
$ grad(O/H)= A+B*\log(M_{halo}/M_*)$&$-380\pm30$&$-320\pm80$&0.60&yes\\
\hline 
\end{tabular}
\end{table*}

\subsection{The relation between the dark halo mass fraction and O/H radial gradient}
Since a halo mass influence  the chemical evolution of gas in a disk, we  checked whether there is a connection between the dark halo mass fraction and the radial distribution of metallicity. For this purpose we used a larger statistics  than we did above, taking the sample of galaxies with available measurements of metallicity gradients from \citet{Pilyugin2014} and the dark halo masses from \citet{SaburovaDelPopolo}.  Note that the metallicity gradient is less sensitive to the calibration uncertainty than the values of O/H.

\begin{figure}[h!]
\includegraphics[width=9.0cm]{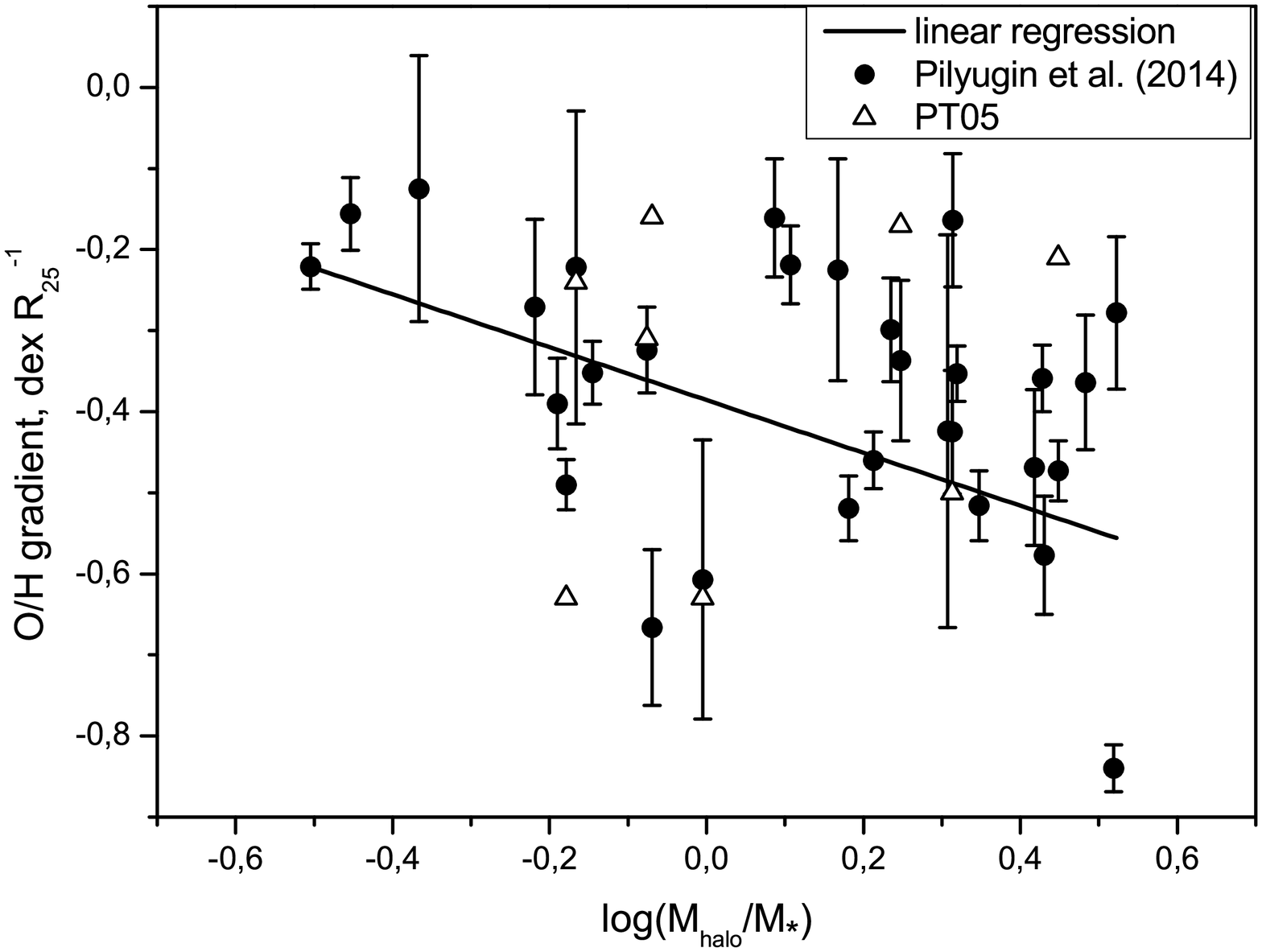}
\caption{The radial gradient of the oxygen abundance taken from \citet{Pilyugin2014} (filled circles) and PT2005 (open triangles) compared to the dark halo to stellar mass ratio within optical borders \citep{SaburovaDelPopolo}. The line denotes the linear regression. Three outliers  with low gradients are NGC3521, NGC2841, NGC5457.}
\label{mhmstarsgradoh}
\end{figure}

In Fig. \ref{mhmstarsgradoh} a mass of dark halo within the optical border,  normalized to the stellar mass of a galaxy $M_*$, is compared with the O/H radial gradient expressed  in terms of dex $R_{25}^{-1}$.  Stellar masses were calculated from the B-band luminosities and $(B-V)_0$ color indices through the \cite{bdj} $M/L$-color model relation.   For comparison,  by open triangles we also show the gradients taken from PT2005 calibration which were used above in the current paper. The statistics is too poor for them to reveal any correlation, however the more representative  \citet{Pilyugin2014} data reveal a shallower gradient for galaxies with low dark halo masses. Three outliers  with low gradients are NGC3521, NGC2841 and NGC5457, which also have the higher central values of O/H than the average of the sample. Two of these galaxies have decreasing radial profiles of the effective yield (see Section 2). So their high gradient of O/H could be the result of the overabundance of oxygen in their inner parts. 
 It is remarkable that if we consider the central value of O/H instead of the gradient, a correlation vanishes, hence it is not of a local nature. Although the link between the halo mass and the (O/H) gradient is not so tight,  one should have in mind that  typical errors of estimations  are comparable with the dispersion of points on the diagram. 

A possible explanation of the correlation is the most efficient radial gas mixing in galaxies with low relative mass of a dark halo 
as the result of  a strong interaction experienced by these galaxies in the past. Interaction or merging of galaxies  leads not only to flattening of the abundance gradient (\citealt{Werk2011}) but also to the partial destruction of their halos (see simulation results of \citealt{Libeskind2011}). In addition, a gas mixing leading to a shallow abundance gradient may also be caused by the turbulent viscosity or accretion of slowly rotating gas (see discussion in \citealt{Elmegreen2014}), although it is not evident that these processes are  more active for low mass halos.
\section{Conclusions}
We considered  the radial profiles of the effective oxygen yield  $y_{eff}$, which corresponds to the true stellar yield  $y_o$ for the closed box model. In real galaxies it may be lower than $y_o$ in the case of accretion or gas outflow from a disk. As the initial data for gas-phase abundance we used the O/H radial profiles for 14 spiral galaxies taken from \citet{Moustakas2010}, which are based on different calibrations (PT2005 \citet{PT2005} and KK2004 \citet{KK2004} methods).  The absolute values and shapes of radial profiles $y_{eff}(R)$ of the effective yield are found to be very different for  galaxies even of similar types, reflecting their different history of evolution. The yield $y_{eff}$ varies with galactocentric distance $R$ within a factor 2 - 3.  For galaxies with prominent bulges both calibrations reveal a local maximum $y_{eff}$(0) which may be interpreted as the result of the enriched gas inflow from bulge onto a disk in the bulge-dominating regions.  In most of galaxies  $y_{eff}$(R)  increases  or remains roughly constant along the radius  for the PT2005 version of O/H (which we consider as a preferable one), remaining  significantly lower than the existing estimates of the true stellar $y_o$.  For KK2004 data the situation is more ambiguous – the effective yield $y_{eff}$(R) grows in 8 out of 14 galaxies. 
 We argue that the relatively low  effective yield  evidences the  accretion of low abundant gas, and, as the radial profiles of $y_{eff}$ show, in many cases a gas inflow is most effective for the inner parts of disks.    A positive correlation between the $y_{eff}$  and  gas mass fraction indicates that the accretion more significantly decreases the effective yield in galaxies with lower gas content. 
 It is essential that the maximal values of the effective yield $y_{eff~max}$ of galaxies  based on the PT2005 data  anti-correlate with the masses of galaxies within the optical radius and with their rotation velocities, being systematically lower for massive galaxies with higher dark halo masses. It demonstrates that the evolution of massive spiral galaxies possessing massive halos is especially different from that expected for a closed box scenario. These results agree with the evolutionary models where the accretion of metal-poor halo gas 
 most effectively reduces the oxygen abundance in the rapidly rotating galaxies with massive halos. We also found that galaxies with higher dark halo-to-stellar mass ratios have a tendency to possess more shallow abundance gradient, although this result requires further verification.

\acknowledgments
We thank the referee, Rolf-Peter Kudritzki for his valuable comments that helped us to improve the paper.
This work was supported by Russian Foundation for Basic Research project No 14-22-03006-ofi-m. AS is grateful to the Russian Foundation for Basic Research projects Nos. 15-32-21062 a, 15-52-15050 and the Russian President's  grant No. MD-7355.2015.2.
\bibliographystyle{aj}
\bibliography{yeff}

\begin{thebibliography}{41}
\providecommand{\natexlab}[1]{#1}
\providecommand{\url}[1]{\texttt{#1}}
\providecommand{\urlprefix}{URL }
\providecommand{\eprint}[2][]{\url{#2}}

\bibitem[{{Ascasibar}, {Gavil{\'a}n}, {Pinto} et~al.(2015)}]{Ascasibar2015}
{Ascasibar}, Y., {Gavil{\'a}n}, M., {Pinto}, N., et~al. 2015, \mnras, 448,
  2126. \eprint{1406.6397}

\bibitem[{{Belfiore}, {Maiolino}, \& {Bothwell}(2015)}]{Belfiore2015}
{Belfiore}, F., {Maiolino}, R., \& {Bothwell}, M. 2015, ArXiv e-prints.
  \eprint{1503.06823}

\bibitem[{{Bell} \& {de Jong}(2001)}]{bdj}
{Bell}, E.~F., \& {de Jong}, R.~S. 2001, \apj, 550, 212.
  \eprint{arXiv:astro-ph/0011493}

\bibitem[{{Bouch{\'e}}, {Dekel}, {Genzel} et~al.(2010)}]{Bouche2010}
{Bouch{\'e}}, N., {Dekel}, A., {Genzel}, R., et~al. 2010, \apj, 718, 1001.
  \eprint{0912.1858}

\bibitem[{{Dalcanton}(2007)}]{Dalcanton2007}
{Dalcanton}, J.~J. 2007, \apj, 658, 941. \eprint{astro-ph/0608590}

\bibitem[{{Edmunds}(1990)}]{Edmunds1990}
{Edmunds}, M.~G. 1990, \mnras, 246, 678

\bibitem[{{Elmegreen}, {Struck}, \& {Hunter}(2014)}]{Elmegreen2014}
{Elmegreen}, B.~G., {Struck}, C., \& {Hunter}, D.~A. 2014, \apj, 796, 110.
  \eprint{1411.0332}

\bibitem[{{Fraternali} \& {Binney}(2006)}]{Fraternali2006}
{Fraternali}, F., \& {Binney}, J.~J. 2006, \mnras, 366, 449.
  \eprint{astro-ph/0511334}

\bibitem[{{Gavil{\'a}n}, {Ascasibar}, {Moll{\'a}} et~al.(2013)}]{Gavilan2013}
{Gavil{\'a}n}, M., {Ascasibar}, Y., {Moll{\'a}}, M., et~al. 2013, \mnras, 434,
  2491. \eprint{1306.6565}

\bibitem[{{Kewley} \& {Ellison}(2008)}]{Kewley2008}
{Kewley}, L.~J., \& {Ellison}, S.~L. 2008, \apj, 681, 1183. \eprint{0801.1849}

\bibitem[{{Kobulnicky} \& {Kewley}(2004)}]{KK2004}
{Kobulnicky}, H.~A., \& {Kewley}, L.~J. 2004, \apj, 617, 240.
  \eprint{astro-ph/0408128}

\bibitem[{{Kudritzki}, {Ho}, {Schruba} et~al.(2015)}]{Kudritzki2015}
{Kudritzki}, R.-P., {Ho}, I.-T., {Schruba}, A., et~al. 2015, \mnras, 450, 342.
  \eprint{1503.01503}

\bibitem[{{Leroy}, {Walter}, {Brinks} et~al.(2008)}]{Leroy2008}
{Leroy}, A.~K., {Walter}, F., {Brinks}, E., et~al. 2008, \aj, 136, 2782.
  \eprint{0810.2556}

\bibitem[{{Libeskind}, {Knebe}, {Hoffman} et~al.(2011)}]{Libeskind2011}
{Libeskind}, N.~I., {Knebe}, A., {Hoffman}, Y., et~al. 2011, \mnras, 418, 336.
  \eprint{1107.4366}

\bibitem[{{Lu}, {Mo}, \& {Wechsler}(2015)}]{Lu2015}
{Lu}, Y., {Mo}, H.~J., \& {Wechsler}, R.~H. 2015, \mnras, 446, 1907.
  \eprint{1402.2036}

\bibitem[{{Mart{\'{\i}}n-Navarro}, {La Barbera}, {Vazdekis}
  et~al.(2014)}]{2014arXiv14046533M}
{Mart{\'{\i}}n-Navarro}, I., {La Barbera}, F., {Vazdekis}, A., et~al. 2014,
  ArXiv e-prints. \eprint{1404.6533}

\bibitem[{{Martinsson}, {Verheijen}, {Westfall} et~al.(2013)}]{Martinsson2013}
{Martinsson}, T.~P.~K., {Verheijen}, M.~A.~W., {Westfall}, K.~B., et~al. 2013,
  \aap, 557, A131. \eprint{1308.0336}

\bibitem[{{Moran}, {Heckman}, {Kauffmann} et~al.(2012)}]{Moran2012}
{Moran}, S.~M., {Heckman}, T.~M., {Kauffmann}, G., et~al. 2012, \apj, 745, 66.
  \eprint{1112.1084}

\bibitem[{{Moustakas}, {Kennicutt}, {Tremonti} et~al.(2010)}]{Moustakas2010}
{Moustakas}, J., {Kennicutt}, R.~C., Jr., {Tremonti}, C.~A., et~al. 2010,
  \apjs, 190, 233. \eprint{1007.4547}

\bibitem[{{Oppenheimer}, {Dav{\'e}}, {Kere{\v s}}
  et~al.(2010)}]{Oppenheimer2010}
{Oppenheimer}, B.~D., {Dav{\'e}}, R., {Kere{\v s}}, D., et~al. 2010, \mnras,
  406, 2325. \eprint{0912.0519}

\bibitem[{{Pilyugin}(2003)}]{Pilyugin2003}
{Pilyugin}, L.~S. 2003, \aap, 399, 1003. \eprint{astro-ph/0211319}

\bibitem[{{Pilyugin}, {Grebel}, \& {Kniazev}(2014)}]{Pilyugin2014}
{Pilyugin}, L.~S., {Grebel}, E.~K., \& {Kniazev}, A.~Y. 2014, \aj, 147, 131.
  \eprint{1403.5461}

\bibitem[{{Pilyugin}, {Grebel}, {Zinchenko} et~al.(2014)}]{Pilyugin2014b}
{Pilyugin}, L.~S., {Grebel}, E.~K., {Zinchenko}, I.~A., et~al. 2014, \aj, 148,
  134. \eprint{1408.6413}

\bibitem[{{Pilyugin}, {Lara-L{\'o}pez}, {Grebel} et~al.(2013)}]{Pilyugin2013}
{Pilyugin}, L.~S., {Lara-L{\'o}pez}, M.~A., {Grebel}, E.~K., et~al. 2013,
  \mnras, 432, 1217. \eprint{1304.0191}

\bibitem[{{Pilyugin} \& {Thuan}(2005)}]{PT2005}
{Pilyugin}, L.~S., \& {Thuan}, T.~X. 2005, \apj, 631, 231

\bibitem[{{Pilyugin}, {Thuan}, \& {V{\'{\i}}lchez}(2007)}]{Pilyugin2007}
{Pilyugin}, L.~S., {Thuan}, T.~X., \& {V{\'{\i}}lchez}, J.~M. 2007, \mnras,
  376, 353. \eprint{astro-ph/0701332}

\bibitem[{{Pilyugin}, {V{\'{\i}}lchez}, \& {Contini}(2004)}]{Pilyugin2004}
{Pilyugin}, L.~S., {V{\'{\i}}lchez}, J.~M., \& {Contini}, T. 2004, \aap, 425,
  849. \eprint{astro-ph/0407014}

\bibitem[{{Reyes}, {Mandelbaum}, {Gunn} et~al.(2012)}]{Reyes2012}
{Reyes}, R., {Mandelbaum}, R., {Gunn}, J.~E., et~al. 2012, \mnras, 425, 2610

\bibitem[{{Rosales-Ortega}, {S{\'a}nchez}, {Iglesias-P{\'a}ramo}
  et~al.(2012)}]{Rosales-Ortega2012}
{Rosales-Ortega}, F.~F., {S{\'a}nchez}, S.~F., {Iglesias-P{\'a}ramo}, J.,
  et~al. 2012, \apjl, 756, L31. \eprint{1207.6216}

\bibitem[{{Saburova} \& {Del Popolo}(2014)}]{SaburovaDelPopolo}
{Saburova}, A., \& {Del Popolo}, A. 2014, \mnras, 445, 3512. \eprint{1410.3052}

\bibitem[{{Saburova} \& {Zasov}(2012)}]{SaburovaZasov}
{Saburova}, A.~S., \& {Zasov}, A.~V. 2012, Astronomy Letters, 38, 139

\bibitem[{{S{\'a}nchez}, {Rosales-Ortega}, {Iglesias-P{\'a}ramo}
  et~al.(2014)}]{Sanchez2014}
{S{\'a}nchez}, S.~F., {Rosales-Ortega}, F.~F., {Iglesias-P{\'a}ramo}, J.,
  et~al. 2014, \aap, 563, A49. \eprint{1311.7052}

\bibitem[{{S{\'a}nchez}, {Rosales-Ortega}, {Jungwiert}
  et~al.(2013)}]{Sanchez2013}
{S{\'a}nchez}, S.~F., {Rosales-Ortega}, F.~F., {Jungwiert}, B., et~al. 2013,
  \aap, 554, A58. \eprint{1304.2158}

\bibitem[{{S{\'a}nchez Almeida}, {Elmegreen}, {Mu{\~n}oz-Tu{\~n}{\'o}n}
  et~al.(2014)}]{Almeida2014}
{S{\'a}nchez Almeida}, J., {Elmegreen}, B.~G., {Mu{\~n}oz-Tu{\~n}{\'o}n}, C.,
  et~al. 2014, \aapr, 22, 71. \eprint{1405.3178}

\bibitem[{{Searle} \& {Sargent}(1972)}]{Searle1972}
{Searle}, L., \& {Sargent}, W.~L.~W. 1972, \apj, 173, 25

\bibitem[{{Spitoni}, {Calura}, {Matteucci} et~al.(2010)}]{Spitoni2010}
{Spitoni}, E., {Calura}, F., {Matteucci}, F., et~al. 2010, \aap, 514, A73.
  \eprint{1001.4374}

\bibitem[{{van de Voort}, {Schaye}, {Booth} et~al.(2011)}]{vandeVoort2011}
{van de Voort}, F., {Schaye}, J., {Booth}, C.~M., et~al. 2011, \mnras, 414,
  2458. \eprint{1011.2491}

\bibitem[{{Vincenzo}, {Matteucci}, {Belfiore} et~al.(2015)}]{Vincenzo2015}
{Vincenzo}, F., {Matteucci}, F., {Belfiore}, F., et~al. 2015, ArXiv e-prints.
  \eprint{1503.08300}

\bibitem[{{Werk}, {Putman}, {Meurer} et~al.(2011)}]{Werk2011}
{Werk}, J.~K., {Putman}, M.~E., {Meurer}, G.~R., et~al. 2011, \apj, 735, 71.
  \eprint{1104.3897}

\bibitem[{{Zahid}, {Dima}, {Kudritzki} et~al.(2014)}]{Zahid2014}
{Zahid}, H.~J., {Dima}, G.~I., {Kudritzki}, R.-P., et~al. 2014, \apj, 791, 130.
  \eprint{1404.7526}

\bibitem[{{Zasov}, {Saburova}, {Katkov} et~al.(2015)}]{Zasov2015}
{Zasov}, A., {Saburova}, A., {Katkov}, I., et~al. 2015, \mnras, 449, 1605.
  \eprint{1503.05220}

\end{thebibliography}

\end{document}